\documentstyle[prc,epsf,aps]{revtex}
\begin{document}

\title{The relativistic eikonal approximation in high-energy $A(e,e'p)$ reactions}
\author{D.\ Debruyne, J.\ Ryckebusch, W.\ Van Nespen and S.\ Janssen}
\address{Department of Subatomic and Radiation Physics, \protect\\
Ghent University, Proeftuinstraat 86, B-9000 Gent, Belgium}
\date{\today}
\maketitle

\begin{abstract}
A fully relativistic model for the description of exclusive $(e,e'p)$
reactions off nuclear targets at high energies and momentum transfers
is outlined.  It is based on the eikonal approximation for the
ejectile scattering wave function and a relativistic mean-field
approximation to the Walecka model.  Results for $^{12}$C$(e,e'p)$ and
$^{16}$O$(e,e'p)$ differential cross sections and separated structure
functions are presented for four-momenta in the range $0.8 \leq Q^{2}
\leq 20 $ (GeV/c)$^{2}$.  The regions of applicability of the eikonal
approximation are studied and observed to be confined to proton
knockout in a relatively small cone about the momentum transfer.  A
simple criterium defining the boundaries of this cone is
determined. The $Q^2$ dependence of the effect of off-shell
ambiguities on the different $(e,e'p)$ structure functions is
addressed.  At sufficiently high values of $Q^2$ their impact on the
cross sections is illustrated to become practically negligible.  It is
pointed out that for the whole range of $Q^2$ values studied here, the
bulk of the relativistic effects arising from the coupling between the
lower components in the wave functions, is manifesting itself in the
longitudinal-transverse interference term.
\end{abstract}

\vspace{0.5cm}


\section{Introduction}

Exclusive A$(e,e'p)$B reactions from nuclei constitute an invaluable
tool to probe a wide variety of nuclear phenomena.  At low values of
the virtual photon's four-momentum transfer $Q^{2}=\vec{q}^2-\omega^2$
and, accordingly, larger distance scales, the quasi-elastic A$(e,e'p)$
reaction probes the mean-field structure of nuclei. From systematic
investigations for a large number of target nuclei a richness of
precise information about the independent-particle wave functions and
spectroscopic strengths was assembled \cite{vijay}. At high $Q^{2}$
and decreasing distance scales, the scope of exclusive $(e,e'p)$
measurements shifts towards studies of (possible) medium dependencies
of the nucleonic properties, and, effects like color transparency and
the short-range structure of nuclei. Within the context of exclusive
$(e,e'p)$ reactions, ``color transparency'' stands for the suggestion
that at sufficiently high values of $Q^2$ the struck proton may interact
in an anomalously weak manner with the ``spectator'' nucleons in the
target nucleus \cite{frankcolor}.

The extraction of physical information from measured A$(e,e'p)$B cross
sections usually involves some theoretical modelling of which the
major ingredients are the initial (bound) and final (scattering)
proton wave functions and the electromagnetic electron-nucleus
coupling. At lower values of $Q^2$, most theoretical work on $(e,e'p)$
reactions was performed in the so-called distorted-wave impulse
approximation (DWIA).  The idea behind the DWIA approach is that the
inital (bound) and final (scattering) state of the struck nucleon can
be computed in a potential model, whereas for the electron-nucleus
coupling an ``off-shell corrected'' electron-proton form can be used.
The wealth of high-quality $(e,e'p)$ data that electron-scattering
experiments have provided over the last 20 years, made sure that the
DWIA models are well tested against experimental data. For higher
values of the energy and momentum transfer ($Q^2 \gtrsim
1~$(GeV/c)$^2$), most theoretical $(e,e'p)$ work starts from the
non-relativistic Glauber theory \cite{jeschonnek}.  This theory is
highly successful in describing small angle proton-nucleus scattering
at higher energies \cite{alkhazov78} and is conceived as a baseline
for calculating the effect of final-state interactions in high-energy
$(e,e'p)$ reactions.  Glauber theory is a multiple-scattering
extension of the standard eikonal approximation that relates through a
profile function the ejectile's distorted wave function to the elastic
proton scattering wave function
\cite{jeschonnek,frankglauber,benhar96,nikolaev96,ciofi99}.  The
Glauber method has frequently been shown to be reliable in describing
A$(p,p')$ processes. Several non-relativistic studies
\cite{radici2,rinat97,seki00} have formally investigated the
applicability of the Glauber model for describing A$(e,e'p)$ reactions
at higher energies and momentum transfers.  Recently, the first
high-quality data for exclusive $^{16}$O$(e,e'p)$ cross sections at
higher four-momentum transfer ($Q^2 \geq 1$(GeV/c)$^2$ became
available \cite{gao}.  Below, we will compare results of relativistic
eikonal calculations with these data.  We believe that this comparison
between model calculations and data provides a stringent test of
the applicability of the eikonal approximation in describing $(e,e'p)$
reactions..

Since relativistic effects are expected to become critical in the GeV
energy domain, we explore the possibility of developing a fully
relativistic model for A$(e,e'p)$ processes, thereby using the eikonal
limit to solve the equations for the final-state wave functions.  We
employ a relativistic mean-field approximation to the Walecka model to
determine the bound state wave functions and binding energies, as well
as nucleon and meson potentials.  The same mean-field potentials are
then also used to compute the scattering wave function in the Dirac
eikonal limit.  The work presented here is a small initial step
towards the formulation of a fully microscopic relativistic model for
the description of $(e,e'p)$ reactions that could possibly bridge the
gap between the low and intermediate-energy regime.  The model
developed in this work can be formally applied in a wide $Q^2$ range.
As a matter of fact, we employ the relativistic eikonal method to
estimate the sensitivity of $(e,e'p)$ observables in the few GeV
regime to a number of physical effects, including off-shell
ambiguities and relativity.  We adopt different prescriptions for the
electron-nucleus coupling in our calculations.  By doing this, we
estimate the sensitivity of the observables to the theoretical
uncertainties that surround the choice of the off-shell
electron-proton vertex.  It is often claimed that off-shell
ambiguities decrease in importance as the four-momentum transfer
increases.  Here, we make an attempt to quantify the relative
importance of the off-shell effects for the $(e,e'p)$ structure
functions by comparing results obtained with different off-shell
electron-proton couplings.  Hereby we are primarily concerned with the
question how big the uncertainties remain when higher and higher
four-momentum transfers are probed.

In Sect.\ \ref{sec:formalism} we introduce a relativistic eikonal
formalism for calculating A$(e,e'p)$ observables.  This includes a
discussion of the method employed to determine the bound (Sect.\
\ref{subsec:bound}) and scattering (Sect.\ \ref{subsec:eikonal})
states.  Various forms for the photon-nucleus interaction vertex are
introduced in Sect.\ \ref{subsec:vertex}, where special attention is
paid to the issue of current conservation.  In Sect.\ \ref{sec:results}
we present the results of our $^{12}$C$(e,e'p)$ and $^{16}$O$(e,e'p)$
numerical calculations.    In Sect.\ \ref{subsec:current} we focus on the issue of
the $Q^2$ evolution of the off-shell ambiguities.  In Sect.\
\ref{subsec:relativity} we compare the results of a fully relativistic
calculation with a calculation in which the explicit coupling between the lower
components in the inital and final state are neglected.  Finally, our
concluding remarks are summarized in Sect.\ \ref{sec:conclusion}.

\section{Formalism}
\label{sec:formalism}

\subsection{Reaction observables and kinematics.}

In this work we follow the conventions for the
$(\vec{e},e'\vec{p})$ kinematics and observables introduced by
Donnelly and Raskin in Ref.\ \cite{donnelly}.  The four-momenta of the
incident and scattered electrons are labeled as $K^{\mu}
(\epsilon,\vec{k})$ and $K^{'\mu} (\epsilon',\vec{k'})$.  The electron
momenta $\vec{k}$ and $\vec{k'}$ define the scattering plane.  The
four-momentum transfer is given by $q^{\mu} = K^{\mu} - K^{'\mu} =
P_{A-1}^{\mu} + P_{f}^{\mu} - P_{A}^{\mu}$, where $P_{A}^{\mu}$ and
$P_{A-1}^{\mu}$ are the four-momenta of the target and residual nucleus,
while $P_{f}^{\mu}$ is the four-momentum of the ejected nucleon.
Also, $q^{\mu} = (\omega,\vec{q})$, where the three-momentum transfer
$\vec{q} = \vec{k} - \vec{k'} = \vec{k}_{A-1} + \vec{k}_{f} -
\vec{k}_{A}$ and the energy transfer $\omega = \epsilon - \epsilon' =
E_{A-1} + E_{f} - E_{A}$ are defined in the standard manner.  The $xyz$
coordinate system is chosen such that the $z$-axis lies along the
momentum transfer $\vec{q}$, the $y$-axis lies along $\vec{k} \times
\vec{k'}$ and the $x$-axis lies in the scattering plane; the reaction
plane is then defined by $\vec{k}_{f}$ and $\vec{q}$.  The
Bjorken-Drell convention \cite{bjorken} for the gamma matrices and Dirac
spinors is followed, so that the normalization condition for Dirac
plane waves, characterized by a four-momentum $K^{\mu}$ and spin-state
$S^{\mu}$, is $\bar{u}(K^{\mu},S^{\mu})u(K^{\mu},S^{\mu}) = 1$.

In the one-photon-exchange approximation, the process in which a
longitudinally polarized electron with helicity h, impinges on a
nucleus and induces the knockout of a single nucleon, leaving the
residual nucleus in a certain discrete state, can be written in the
following form \cite{donnelly} :
\begin{eqnarray}
\frac{d^{5}\sigma}{d\epsilon'd\Omega_{e'}d\Omega_{x}} & = &
\frac{MM_{A-1}k_{f}}{8\pi^{3}M_{A}} f_{rec}^{-1} \sigma_{M}
\biggl[(v_{L}{\cal R}_{L}+v_{T}{\cal R}_{T}+v_{TT}{\cal
R}_{TT}+v_{TL}{\cal R}_{TL}) \nonumber \\ & + &
h(v_{T'}{\cal R}_{T'}+v_{TL'}{\cal R}_{TL'})\biggr] \; ,
\end{eqnarray}  
where $f_{rec}$ is the hadronic recoil factor
\begin{eqnarray}
f_{rec} = \frac{E_{A-1}}{E_{A}} \left| 1+\frac{E_{f}}{E_{A-1}}
\left(1-\frac{\vec{q} \cdot \vec{k}_{f}}{k_{f}^{2}} \right)\right| =
\left| 1+\frac{\omega k_{f} - E_{f} q \cos\theta_{f}}{M_{A}k_{f}}
\right| \; ,
\end{eqnarray}
with $\theta_{f}$ the angle between $\vec{k_{f}}$ and $\vec{q}$, and
$\sigma_{M}$ the Mott cross section 
\begin{eqnarray}
\sigma_{M} = \left( \frac{\alpha\cos\theta_{e}/2}{2\epsilon\sin^{2}\theta_{e}/2}\right)^{2} \; ,
\end{eqnarray} 
with $\theta_{e}$ the angle between the incident and the scattered electron.
The electron kinematics is contained in the kinematical factors
\begin{eqnarray}
v_{L} & = & \left(\frac{Q^{2}}{q^{2}}\right)^{2} \; , \\ v_{T} & = &
-\frac{1}{2} \left(\frac{Q^{2}}{q^{2}}\right) + \tan^{2}
\frac{\theta_{e}}{2} \; , \\ v_{TT} & = & \frac{1}{2}
\left(\frac{Q^{2}}{q^{2}}\right) \; , \\ v_{TL} & = &
\frac{1}{\sqrt{2}} \left(\frac{Q^{2}}{q^{2}}\right)
\sqrt{-\left(\frac{Q^{2}}{q^{2}}\right)+ \tan^{2}
\frac{\theta_{e}}{2}} \; , \\ v_{T'} & = & \tan \frac{\theta_{e}}{2}
\sqrt{-\left(\frac{Q^{2}}{q^{2}}\right)+ \tan^{2}
\frac{\theta_{e}}{2}} \; , \\ v_{TL'} & = & \frac{1}{\sqrt{2}}
\left(\frac{Q^{2}}{q^{2}}\right) \tan \frac{\theta_{e}}{2} \; ,
\end{eqnarray} 
whereas the structure functions are defined in a standard fashion as
\begin{eqnarray}
{\cal R}_{L} & = & |\rho(\vec{q})_{fi}|^{2} \; , \\ {\cal R}_{T} & = &
|J(\vec{q};+1)_{fi}|^{2} + |J(\vec{q};-1)_{fi}|^{2}\; , \\ {\cal
R}_{TT} & = & 2 \, {\textstyle Re} \,
\{J^{\star}(\vec{q};+1)_{fi}J(\vec{q};-1)_{fi}\} \; , \\ {\cal R}_{TL}
& = & -2 \, {\textstyle Re} \, \{\rho^{\star}(\vec{q})_{fi}
(J(\vec{q};+1)_{fi}-J(\vec{q};-1))_{fi}\} \; , \\ {\cal R}_{T'} & = &
|J(\vec{q};+1)_{fi}|^{2} - |J(\vec{q};-1)_{fi}|^{2}\; , \\ {\cal
R}_{TL'} & = & -2 \, {\textstyle Re} \, \{\rho^{\star}(\vec{q})_{fi}
(J(\vec{q};+1)_{fi}+J(\vec{q};-1))_{fi}\} \; ,
\end{eqnarray}
where $\rho(\vec{q})_{fi}$ is the transition charge density, while
$J(\vec{q};m=\pm1)_{fi}$ is the transition three-current expanded in
terms of the standard spherical components.  

\subsection{Bound state wave functions.}
\label{subsec:bound}

A relativistic quantum field theory for nucleons ($\psi$) interacting
with scalar mesons ($\phi$) through a Yukawa coupling $\bar{\psi} \psi
\phi$ and with neutral vector mesons ($V_{\mu}$) that couple to the
conserved baryon current $\bar{\psi} \gamma_{\mu} \psi$, can be
described through a lagrangian density of the type
\cite{walecka,horser}
\begin{eqnarray}
{\cal L}_{0}  & = &  
\bar{\psi} ( \imath \!\! \not{\partial} - M) \psi 
+ \frac{1}{2} ( \partial_{\mu} \phi \partial^{\mu} \phi - m_{s}^{2} \phi^{2} ) 
- \frac{1}{4} G_{\mu\nu} G^{\mu\nu} 
\nonumber \\
& & 
+ \frac{1}{2} m_{v}^{2} V_{\mu} V^{\mu} 
- g_{v} \bar{\psi} \gamma_{\mu} \psi V^{\mu} 
+ g_{s} \bar{\psi} \psi \phi 
\; ,
\end{eqnarray}
with $M$, $m_s$ and $m_v$ the nucleon, scalar meson and vector meson
masses, respectively, and $G^{\mu\nu} \equiv \partial^{\mu} V^{\nu} -
\partial^{\nu} V^{\mu}$ the vector meson field strength.  The scalar
and vector fields may be associated with the $\sigma$ and $\omega$
mesons.  The model can be extended to include also $\pi$ and
$\rho$ mesons, as well as the coupling to the photon field.  The
corresponding lagrangian has the form
\begin{eqnarray} 
{\cal L} & = & {\cal L}_{0} + \frac{1}{2} ( \partial_{\mu} \vec{\pi}
\cdot \partial^{\mu} \vec{\pi} - m_{\pi}^{2} \vec{\pi} \cdot \vec{\pi}
) - \imath g_{\pi} \bar{\psi} \gamma_{5} \vec{\tau} \cdot \vec{\pi}
\psi - \frac{1}{4} \vec{B}_{\mu\nu} \cdot \vec{B}^{\mu\nu} \nonumber
\\ & & + \frac{1}{2} m_{\rho}^{2} \vec{b}_{\mu} \cdot \vec{b}^{\mu} -
\frac{1}{2} g_{\rho} \bar{\psi} \gamma_{\mu} \vec{\tau} \cdot
\vec{b}^{\mu} \psi - \frac{1}{4} F_{\mu\nu} F^{\mu\nu} \nonumber \\ &
& - e A_{\mu} [ \bar{\psi} \gamma^{\mu} \frac{1}{2} ( 1 + \tau_{3} )
\psi + ( \vec{b}_{\nu} \times \vec{B}^{\nu\mu} )_{3} + ( \vec{\pi}
\times ( \partial^{\mu} \vec{\pi} + g_{\rho} ( \vec{\pi} \times
\vec{b}^{\mu} ) ) )_{3} ] \; .
\label{eq:fullan}
\end{eqnarray}
Here $\vec{\pi}$, $\vec{b}_{\mu}$, $A_{\mu}$, $F_{\mu\nu}$ are the
pion, rho, Maxwell and electromagnetic fields. Further,
$\vec{B}^{\mu\nu} \equiv \partial^{\mu} \vec{b}^{\nu} - \partial^{\nu}
\vec{b}^{\mu} - g_{\rho} ( \vec{b}^{\mu} \times \vec{b}^{\nu} )$ is
the $\rho$-meson field.

At sufficiently high densities, the meson field operators can be
approximated by their expectation values.  Within the context of the
relativistic Hartree approximation, it can be shown that when starting
from the langrangian (\ref{eq:fullan}) the following Dirac equation
for the baryon field $\Psi$ results \cite{horser} :
\begin{eqnarray}
\left[ \imath \gamma^{\mu} \partial_{\mu} - M - \Sigma_{H} \right]
\Psi = 0 \; ,
\end{eqnarray}
where the self-energy $\Sigma_{H}$ is defined as
\begin{eqnarray}
\Sigma_{H}  =  - g_{s} \phi + g_{v} \gamma_{\mu} V^{\mu} + g_{\pi}
\gamma_{5} \tau_{\alpha} \pi^{\alpha} + \frac{1}{2} g_{\rho} \gamma_{\mu}
\tau_{\alpha} b^{\mu\alpha} + \frac{1}{2} \gamma_{\mu} ( 1 + \tau_{3}
) A^{\mu} \; . 
\end{eqnarray}
Assuming that the nuclear ground state is spherically symmetric
and a parity eigenstate, it can be shown that the pion field does not
enter in the Hartree approximation.  
Furthermore, the meson fields only depend on the radius, and only the
time component of the vector fields contribute.
The time-independent Dirac equation can then be written as :
\begin{eqnarray}
\label{eq:hamiltonian}
\hat{H} \Psi(\vec{x}) & \equiv & \left[ - \imath \vec{\alpha} \cdot
\vec{\nabla} + g_{v} V^{0} (r) + \frac{1}{2} g_{\rho} \tau_{\alpha}
b^{0 \alpha} (r) \right. \nonumber \\
& & \left. + \frac{1}{2} e (1+\tau_{3}) A^{0} (r) + \gamma^{0}
(M - g_{s} \phi^{0} (r)) \right] = E \Psi
(\vec{x}) \; .
\end{eqnarray}
The general solutions to a Dirac equation with
spherically symmetric potentials have the form
\begin{equation}
\psi_{\alpha} ( \vec{x} ) \equiv \psi_{n \kappa m t} ( \vec{x} ) = \left[
\begin{array}{c}
\imath G_{n \kappa t} ( r ) / r \; {\cal Y}_{\kappa m} \eta_{t} \\
- F_{n \kappa t} ( r ) / r \; {\cal Y}_{- \kappa m} \eta_{t} 
\end{array} 
\right] \; ,
\end{equation}
where $n$ denotes the principal, $\kappa$ and $m$ the generalized
angular momentum and $t$ the isospin quantum numbers. The ${\cal
Y}_{\pm \kappa m}$ are the well-known spin spherical harmonics and
determine the angular and spin parts of the wavefunction,
\begin{eqnarray}
{\cal Y}_{\kappa m} = \sum_{m_{l}m_{s}}
<lm_{l}\frac{1}{2}m_{s}|l\frac{1}{2}jm> Y_{l,m_{l}} \chi_{\frac{1}{2}
m_{s}} \; , \nonumber \\ 
j = |\kappa| - \frac{1}{2} \; , \hspace{0.8cm} l = \left\{
\begin{array}{ll}
\kappa, & \kappa > 0 \\
-(\kappa+1), & \kappa < 0 \; .
\end{array}
\right.
\end{eqnarray}
The Hartree approximation leads to the following set of
coupled equations for the different fields \cite{horser} :
\begin{eqnarray}
\label{eq:setof}
\frac{d^{2}}{dr^{2}} \phi_{0} ( r ) + \frac{2}{r} \frac{d}{dr}
\phi_{0} ( r ) - m_s^{2} \phi_{0} ( r ) & = & - g_{s} \rho_{s} ( r )
\nonumber \\ & \equiv & - g_{s} \sum_{\alpha_{occ}} \left( \frac{2
\jmath_{\alpha} + 1}{4 \pi r^{2}} \right) ( | G_{\alpha} ( r ) |^{2} -
| F_{\alpha} ( r ) |^{2} ) \; , \nonumber \\ \frac{d^{2}}{dr^{2}}
V_{0} ( r ) + \frac{2}{r} \frac{d}{dr} V_{0} ( r ) - m_v^{2} V_{0} ( r
) & = & - g_{v} \rho_{B} ( r ) \nonumber \\ & \equiv & - g_{v}
\sum_{\alpha_{occ}} \left( \frac{2 \jmath_{\alpha} + 1}{4 \pi r^{2}}
\right) ( | G_{\alpha} ( r ) |^{2} + | F_{\alpha} ( r ) |^{2} ) \; ,
\nonumber \\ \frac{d^{2}}{dr^{2}} b_{0} ( r ) + \frac{2}{r}
\frac{d}{dr} b_{0} ( r ) - m_{\rho}^{2} \phi_{0} ( r ) & = & -
\frac{1}{2} g_{\rho} \rho_{3} ( r ) \nonumber \\ & \equiv & -
\frac{1}{2} g_{\rho} \sum_{\alpha_{occ}} \left( \frac{2
\jmath_{\alpha} + 1}{4 \pi r^{2}} \right) ( | G_{\alpha} ( r ) |^{2} +
| F_{\alpha} ( r ) |^{2} ) ( -1 )^{t_{\alpha} - 1/2} \; , \nonumber \\
\frac{d^{2}}{dr^{2}} A_{0} ( r ) + \frac{2}{r} \frac{d}{dr} A_{0} ( r
) & = & - e \rho_{P} ( r ) \nonumber \\ & \equiv & - e
\sum_{\alpha_{occ}} \left( \frac{2 \jmath_{\alpha} + 1}{4 \pi r^{2}}
\right) ( | G_{\alpha} ( r ) |^{2} + | F_{\alpha} ( r ) |^{2} )
(t_{\alpha} + \frac{1}{2} ) \; , \nonumber \\ \frac{d}{dr} G_{\alpha}
( r ) + \frac{\kappa}{r} G_{\alpha} ( r ) & - & [ \epsilon_{\alpha} -
g_{v} V_{0} ( r ) - t_{\alpha} g_{\rho} b_{0} ( r ) \nonumber \\ & - &
( t_{\alpha} + \frac{1}{2} ) e A_{0} ( r ) + M - g_{s} \phi_{0} ( r )
] F_{\alpha} ( r ) = 0 \; , \nonumber \\ \frac{d}{dr} F_{\alpha} ( r )
- \frac{\kappa}{r} F_{\alpha} ( r ) & + & [ \epsilon_{\alpha} - g_{v}
V_{0} ( r ) - t_{\alpha} g_{\rho} b_{0} ( r ) \nonumber \\ & - & (
t_{\alpha} + \frac{1}{2} ) e A_{0} ( r ) - M + g_{s} \phi_{0} ( r ) ]
G_{\alpha} ( r ) = 0 \; , \nonumber \\ \int_{0}^{\infty} dr \: ( |
G_{\alpha} |^{2} + | F_{\alpha} |^{2} ) & = & 1 \; .
\end{eqnarray}
The above  equations constitute the basis of the relativistic
mean-field approach to the lagrangian of Eq.~(\ref{eq:fullan}).
 
A new computer program to solve the above set of coupled non-linear
differential equations was developed.  Starting from an initial guess
of the Woods-Saxon form for the scalar and vector potential, the Dirac
equations can be solved iteratively using a shooting point method.
Analytic solutions to the equations in the regions of large and small
r allow to impose the proper boundary conditions.  Once the nucleon
wave functions are obtained, the densities and meson fields can be
re-evaluated.  This procedure is repeated a number of times until
convergence for the energy eigenvalues is reached.  We adopt the
values for the $\sigma$, $\omega$ and $\rho$ masses and coupling
constants as they were introduced by Horowitz and Serot \cite{horser}.

For the $^{12}$C and $^{16}$O nuclei, the newly developed C-code {\tt
SOR} performed all integrations for a radial extension of the nucleus
of 20 fm and a stepsize of 0.01 fm.  The coupled Dirac equations were
solved for a shooting point lying at 2 fm using a fourth order
Runge-Kutta algorithm.  As a convergence criterium we imposed a
tolerance level as small as 0.001 MeV on all single-particle energy
levels. The computed densities for the nuclei $^{12}$C and $^{16}$O,
are depicted in Fig.~\ref{fig:densities}.  We have verified that these
results are comparable to those produced by the {\tt TIMORA} code
\cite{horser}, which is widely used to solve the set of Eqs.~(\ref{eq:setof}).

\subsection{The eikonal final state.}
\label{subsec:eikonal}

To construct the scattering states for the ejected nucleons,
we consider the hamiltonian (\ref{eq:hamiltonian}) that was already
used to calculate the bound state wave functions
\begin{eqnarray}
\hat{H} \equiv - \imath \vec{\alpha} \cdot \vec{\nabla} + \gamma^{0} M + \gamma^{0} \Sigma_{H} ( r ) \; ,
\end{eqnarray}
where the self-energy $\Sigma_{H} (r)$ is given by
\begin{eqnarray}
\Sigma_{H} ( r ) = - g_{s} \phi_{0} ( r ) + g_{v} \gamma_{0} V^{0} ( r
) + \frac{1}{2} g_{\rho} \gamma_{0} \tau_{\alpha} b^{0 \alpha} ( r ) +
\frac{1}{2} e \gamma_{0} ( 1 + \tau_{3} ) A^{0} ( r ) \; .
\end{eqnarray}
With the formal substitutions 
\begin{eqnarray}
V_{s} ( r ) & \equiv & - g_{s} \phi_{0} \; , \nonumber \\ V_{v} ( r )
& \equiv & g_{v} V_{0} ( r ) + \frac{1}{2} g_{\rho} b_{0} ( r ) ( - 1
)^{t_{\alpha} - 1/2} + e A_{0} ( r ) (t_{\alpha} +
\frac{1}{2}) \; ,
\end{eqnarray}
the time independent Dirac equation for a projectile with relativistic
energy $E = \sqrt{k^{2} + M^{2}}$ and spin state $s$, can be cast in
the form
\begin{eqnarray}
\hat{H} \phi_{\vec{k},s}^{(+)} = [ \vec{\alpha} \cdot \vec{p}
+ \beta M + \beta V_{s} (r) + V_{v} (r) ] \phi_{\vec{k},s}^{(+)} \; ,
\end{eqnarray}
where we have introduced the notation $\phi_{\vec{k},s}^{(+)}$ for the
unbound Dirac states.
The computed scalar and vector potentials for the $^{12}$C and
$^{16}$O nuclei are displayed in Fig. \ref{fig:potentials}

After some straightforward manipulations, a Schr\"{o}dinger-like
equation for the upper component can be obtained
\begin{eqnarray}
\label{eq:uppereq}
\left[ \frac{p^{2}}{2M} + V_{c} + V_{so} (\vec{\sigma} \cdot \vec{L} -
\imath \vec{r} \cdot \vec{p} ) \right] u_{\vec{k},s}^{(+)} =
\frac{k^{2}}{2M} u_{\vec{k},s}^{(+)} \; ,
\end{eqnarray}
where the central and spin orbit potentials $V_{c}$ and $V_{so}$ are
defined as
\begin{eqnarray}
V_{c} (r) & = & V_{s} (r) + \frac{E}{M} V_{v} (r) + \frac{V_{s}
(r)^{2} - V_{v} (r)^{2}}{2M} \; , \nonumber \\ V_{so} (r) & = &
\frac{1}{2M[E+M+V_{s}(r)-V_{v}(r)]} \frac{1}{r} \frac{d}{dr}
[V_{s}(r)-V_{v}(r)] \; .
\end{eqnarray}
In computing the scattering wave functions, we use the scalar and
vector potentials as obtained from the iterative bound state
calculations.  As a result the initial and final state wave functions
are orthogonalized and no spurious contributions can be expected to
enter the calculated cross sections.

Since the lower component is related to the upper one through
\begin{eqnarray}
w_{\vec{k},s}^{(+)} = \frac{1}{E+M+V_{s}-V_{v}} \vec{\sigma} \cdot
\vec{p} \, u_{\vec{k},s}^{(+)} \; ,
\end{eqnarray}
the solutions to the equation (\ref{eq:uppereq}) determine the
complete relativistic eigenvalue problem.  So far no approximations
have been made.  Various groups \cite{udias,yin,johansson} have solved
the Dirac equation (\ref{eq:uppereq}) for the final scattering state
using Dirac optical potentials derived from global fits to elastic
proton scattering data ~\cite{cooper}.  Not only are global
parametrizations of Dirac optical potentials usually restricted to
proton kinetic energies $T_{p} \leq 1$~GeV, calculations based on
exact solutions of the Dirac equation frequently become impractical at
higher energies.  This is particularly the case for approaches that
rely on partial-wave expansions in determining the transition matrix
elements.  To overcome these complications, we solve the Dirac
equation (\ref{eq:uppereq}) in the eikonal limit \cite{ito,benhar}.
In intermediate-energy proton scattering ($T_p \approx$ 500~MeV) the
eikonal approximation was shown to reproduce fairly well the exact
Dirac partial wave results \cite{amado83}.  Following the discussion of
Ref.~\cite{amado83}, we define the average momentum $\vec{K}$ and the
momentum transfer $\vec{q}$ in terms of the projected initial
($\vec{k}_{i}$) and final momentum ($\vec{k}_{f}$) of the ejectile
\begin{eqnarray}
\vec{q}  & = &  \vec{k}_{f} - \vec{k}_{i} \; , \\
\vec{K}  & = &  \frac{1}{2} (\vec{k}_{f} + \vec{q}) \; .
\end{eqnarray}
In the eikonal, or, equivalently, the small-angle
approximation ($q \gg k_{i}$) the following operatorial substitution
is made in computing the scattering wave function 
\begin{eqnarray}
\label{eq:eiktrans}
p^{2} = [(\vec{p} - \vec{K}) + \vec{K}]^{2} \longrightarrow 2 \vec{K} \cdot
\vec{p} - K^{2} \; .
\end{eqnarray}
After introducing this approximate relation, the Dirac equation for
the upper component (\ref{eq:uppereq}) becomes
\begin{eqnarray}
[- \imath \vec{K} \cdot \vec{\nabla} - K^{2} + M ( V_{c} + V_{so} [ \vec{\sigma}
\cdot ( \vec{r} \times \vec{K} ) - \imath \vec{r} \cdot \vec{K} ])]
u_{\vec{k},s}^{(+)} = 0 \; ,
\label{eq:lineardirac}
\end{eqnarray}
where the momentum operators in the spin orbit and Darwin terms are
substituted by $\vec{K}$.  Remark that the above equation is now linear in
the momentum operator.  In the eikonal limit, the scattering wave
functions take on the form
\begin{eqnarray}
u_{\vec{k},s}^{(+)} = e^{\imath \vec{k} \cdot \vec{r}} e^{\imath
S(\vec{r})} \chi_{\frac{1}{2}m_{s}} \; .
\end{eqnarray}
Inserting this into Eq.\ (\ref{eq:lineardirac}), yields an expression
for the eikonal phase \cite{ito}.  Defining the $z$-axis along the
direction of the average momentum $\vec{K}$, this phase can be written
in an integral form as :
\begin{eqnarray}
\imath S(\vec{b},z) = - \imath \frac{M}{K} \int_{-\infty}^{z} dz' \, [
V_{c} (\vec{b},z') + V_{so} (\vec{b},z') [ \vec{\sigma} \cdot (\vec{b}
\times \vec{K} )- \imath Kz']] \; ,
\label{eq:eikonalphase}
\end{eqnarray}
where we have introduced the notation $\vec{r} \equiv (\vec{b},z)$.
The scattering wave function, which is proportional to
\begin{eqnarray}
\phi_{\vec{k},s}^{(+)} \sim 
\left[
\begin{array}{c}
1 \\
\frac{1}{E+M+V_{s}-V_{v}} \vec{\sigma} \cdot \vec{p} 
\end{array} 
\right]
e^{\imath \vec{k} \cdot \vec{r}} e^{\imath S(\vec{r})} \chi_{\frac{1}{2}m_{s}} \; ,
\end{eqnarray}
is normalized such that
\begin{eqnarray}
\overline{\phi_{\vec{k},s}^{(+)}} \phi_{\vec{k},s}^{(+)} = 1 \; .
\end{eqnarray}
This wave function differs from the plane-wave solution in two
respects.  First, the lower component exhibits the dynamical
enhancement due to the combination of the scalar and vector
potentials.  Second, the eikonal phase $e^{\imath S(\vec{r})}$
accounts for the interactions that the struck nucleon undergoes in its way
out of the nucleus.  The calculation of the eikonal phase
(\ref{eq:eikonalphase}) involves a transformation to a reference frame
other than the usual laboratory or center-of-mass frame, namely the
frame where the average momentum is pointing along the $z$-axis.  As
the eikonal phase has to be re-evaluated for every $(\vec{b},z)$ point
in space, the Dirac eikonal $(e,e'p)$ calculations are very demanding
as far as computing power is concerned.  In evaluating the matrix
elements, the radial integrations were performed on a 0.1 fm mesh.
It is worth remarking that the standard Glauber approach followed in
many studies involves an extra approximation apart from the ones
discussed above.  Indeed, in evaluating the eikonal phase from
Eq.~(\ref{eq:eikonalphase}) one frequently approximates the z-dependence of the
potentials by a delta function.  

\subsection{Off-shell electron-proton coupling}
\label{subsec:vertex}

We express the matrix elements of the nucleon current in the usual form 
\begin{eqnarray}
<P_{f}S_{f}|J^{\mu}|P_{i}S_{i}> = \bar{u}_{f} \Gamma^{\mu}(P_{f},P_{i})u_{i} \; ,
\end{eqnarray}
where $\Gamma^{\mu}$ is the electromagnetic vertex function for the
nucleon and $u_{i}$ ($u_{f}$) the nucleon spinors.  As discussed in
many works \cite{pollock,nagorny,gross,kellygauge,kellypol}, some
arbitrariness, often referred to as the ``off-shell ambiguity'',
surrounds the choice for the functional form of the vertex function
$\Gamma^{\mu}$.  For a free nucleon, $\Gamma^{\mu}$ can be expressed
in several fully equivalent forms
\begin{eqnarray}
\label{eq:cc1}
\Gamma_{cc1}^{\mu} & = & G_{M} (Q^{2}) \gamma^{\mu} -
\frac{\kappa}{2M} F_{2} (Q^{2}) (P_{i}^{\mu}+P_{f}^{\mu}) \; , \\
\label{eq:cc2}
\Gamma_{cc2}^{\mu} & = & F_{1} (Q^{2}) \gamma^{\mu} + \imath
\frac{\kappa}{2M} F_{2} (Q^{2}) \sigma^{\mu\nu} q_{\nu} \; , \\
\label{eq:cc3}
\Gamma_{cc3}^{\mu} & = & \frac{1}{2M} F_{1} (Q^{2})
(P_{i}^{\mu}+P_{f}^{\mu}) + \imath \frac{1}{2M} G_{M} (Q^{2})
\sigma^{\mu\nu} q_{\nu} \; ,
\end{eqnarray}
where $F_{1}$ is the Dirac, $F_{2}$ the Pauli form factor and $\kappa$
is the anomalous magnetic moment.  The relation with the Sachs
electric and magnetic form factors is established through $G_{E} =
F_{1} - \tau \kappa F_{2}$ and $G_{M} = F_{1} + \kappa F_{2}$, with
$\tau \equiv Q^{2}/4m^{2}$.

When considering bound (or, ``off-shell'') nucleons, however, the
above vertex functions can no longer be guaranteed to produce the same
results.  As a matter of fact, explicit current conservation is rather
an exception than a rule in most calculations that deal with $(e,e'p)$
reactions from finite nuclei.  In nuclear physics, the most widely
used procedure to ``effectively'' restore current conservation is
based on modifying the longitudinal component of the nuclear vector
current using the substitution
\begin{eqnarray}
\label{eq:long}
J_{z} \; \rightarrow \; \frac{\omega}{q} J_{0} \; .
\end{eqnarray}
This procedure is partly inspired on the observation that
meson-exchange and isobar terms enter the charge current operator in a
higher relativistic order than they used to do for the vector current.
There exist several other prescriptions which are meant to restore
current conservation.  Along similar lines, the charge operator can be
replaced by
\begin{eqnarray}
\label{eq:charge}
J_{0} \; \rightarrow \; \frac{q}{\omega} J_{z} \; .
\end{eqnarray}
One can also construct a vertex function that garantuees current
conservation for any initial and final nucleon state.  This can be
achieved for example by adding an extra term to the vertex \cite{rpwia}
\begin{eqnarray}
\label{eq:donn}
\Gamma_{DON}^{\mu} & = & F_{1} (Q^{2}) \gamma^{\mu} + \imath
\frac{\kappa}{2M} F_{2} (Q^{2}) \sigma^{\mu\nu} q_{\nu} + F_{1}
(Q^{2}) \frac{\not{q} q^{\mu}}{Q^{2}} \; ,
\end{eqnarray}
which is also equivalent to the Eqs.\ (\ref{eq:cc1}-\ref{eq:cc3}) in
the free nucleon case.  An operator derived from 
the generalized Ward-Takahashi identity 
reads \cite{gross}
\begin{eqnarray}
\label{eq:wt}
\Gamma_{WT}^{\mu}  & = &  \gamma^{\mu} - \imath \frac{\kappa}{2M}
F_{2} (Q^{2}) \sigma^{\mu\nu} q_{\nu} + [F_{1} (Q^{2}) -1] \frac{\not{q}q^{\mu}+Q^{2}\gamma^{\mu}}{Q^{2}} \; .
\end{eqnarray}    

\section{Results}
\label{sec:results}

\subsection{Final state interactions and the eikonal approximation}

We start our $(e,e'p)$ investigations within the relativistic eikonal
approximation for the kinematics of an $^{16}$O$(e,e'p)$ experiment
that was recently performed at Jefferson Lab \cite{gao}.  In this
experiment, the separated $^{16}$O$(e,e'p)$ structure functions are
measured at $Q^{2}$ = 0.8~(GeV/c)$^{2}$ and $\omega$ = 0.439~GeV for
missing (or, initial) proton momenta $p_{m} = \mid \vec{k_{f}} -
\vec{q} \mid $ below 355 MeV/c.  The variation in missing momentum was
achieved by varying the detection angle of the ejected proton with
respect to the direction of the momentum transfer
(``quasi-perpendicular kinematics''). The measured cross sections for
knockout from the $1p_{1/2}$ and $1p_{3/2}$ levels are depicted in
Fig.\ \ref{fig:gaofig} along with the predictions of our relativistic
eikonal calculations.  A spectroscopic factor of 0.6 was adopted for
all bound levels, and the standard dipole form was used for the
electromagnetic form factors.  At low missing momenta, the eikonal
results shown in Fig. \ref{fig:gaofig} produce a fair description of
the data.  As a comparison, the results of a relativistic plane wave
calculation in the impulse approximation (RPWIA) are also displayed.
Through comparing the plane-wave and the eikonal calculations, thereby
keeping all other ingredients of the calculations identical, one can
evaluate how the eikonal method deals with final state interactions
(FSI). In the eikonal calculations, the dips of the RPWIA calculations
are filled in, and, at low missing momenta the RPWIA cross sections
are reduced.  These two features reflect nothing but the usual impact
of the final-state interactions on the A$(e,e'p)$ angular cross
sections.  The limitations of the eikonal approximation ($q \gg
k_{i}$) are immediately visible at higher missing momenta ($p_m \ge$
250~MeV/c).  Here, the eikonal cross sections largely overshoot both
the RPWIA results and the data and should by no means be considered as
realistic.  It is worth remarking that the data closely follow the
trend set by the RPWIA curves.  As a matter of fact, whereas the
eikonal calculations predict huge effects from final-state
interactions at large transverse missing momenta, the data seem to
suggest rather the opposite effect.  We consider this observation as
one of the major findings of this work.  

One may wonder whether the observed behaviour of the eikonal results
at higher missing momenta in Fig.~\ref{fig:gaofig} is a mere
consequence of the small-angle approximation contained in
Eq.~(\ref{eq:eiktrans}), or whether the adopted model assumptions for
computing the scattering states is also (partly) at the origin of this
pathological behavior.  To address this question, we have performed
calculations for various fixed recoil angles $\theta$ defined as
\begin{equation}
\cos \theta = \frac {\vec{p}_{m} \cdot \vec{q} } 
{\left|   \vec{p}_{m} \right| \left| \vec{q} \right| } \; .   
\end{equation}
The results are displayed in terms of the reduced cross section $\rho$
which is defined in the standard fashion as the differential cross
section, divided by a kinematical factor times the $``CC1''$ off-shell
electron-nucleon cross section of Ref.\ \cite{forest}.  For the
results of Figure \ref{fig:fixtp} we considered in-plane kinematics at
a fixed value of the outgoing proton momentum ($k_f$=1~(GeV)/c) and an
initial electron energy of 2.4~GeV. The variation in missing momentum
is achieved by changing the $q$.  For recoil angles $\theta$ = 0$^o$
(``parallel kinematics'') the eikonal calculations do not exhibit an
unrealistic behavior up to $p_m$=0.5~GeV/c, which is the highest
missing momentum considered here.  With increasing recoil angles, and
consequently, growing ``transverse'' components in the missing momenta
the ``unrealistic'' behaviour of the eikonal results becomes manifest.
Accordingly, the accuracy of the eikonal method based on the
small-angle approximation of Eq.\ (\ref{eq:eiktrans}) can only be
guaranteed for proton knockout in a small cone about the momentum
transfer.  A similar quantitative behaviour as a function of the
recoil angle to what is observed in Fig.~\ref{fig:fixtp} was reported
in Ref.\ \cite{jeschonnek} for d$(e,e'p)$n cross sections determined
in a Glauber framework.  We conclude this section with remarking that
the eikonal method does not exclude situations with high initial (or,
missing) momenta, it only requires that the perpendicular component of
ejectiles's momentum $\vec{k}_{f}$ is sufficiently small.  It speaks
for itself that such conditions are best fulfilled as one approaches
parallel kinematics. This observation puts serious constraints on the
applicability of the Glauber method, that is based on the eikonal
approximation, for modelling the final-state interactions in
high-energy $(e,e'p)$ reactions from nuclei.  However, it should be
noted that our framework does use purely real scalar and vector
potentials. More realistic scattering potentials demand an imaginary
part that accounts for the inelastic channels that are open during the
reaction process. The Glauber approach effectively includes these
inelastic channels and on these grounds one may expect that its range
of applicability is somewhat wider than what is observed here.  With
the eye on defining the region of validity for the eikonal
approximation more clearly, we have studied differential cross
sections for various $Q^{2}$.  In Fig.~\ref{fig:cone}, we display the
computed differential cross sections for the
$^{12}$C$(e,e'p)^{11}$B$(1p_{3/2}^{-1})$ process against the missing
momentum for $Q^{2}$ varying between 1 and 20~(GeV/c)$^{2}$.  Hereby,
quasi-elastic conditions were imposed.  The arrow indicates the
missing momentum where the slope of the eikonal differential cross
section starts deviating from the trend set by the RPWIA cross
section.  In the light of the conclusions drawn from the comparison
between data and the eikonal curves in Fig.\ \ref{fig:gaofig}, the
eikonal results should be regarded with care beyond this missing
momentum.  Furthermore, it is clear that the change in the slope of
the angular cross section becomes more and more pronounced as $Q^{2}$
increases.  It is apparent from Fig.\ \ref{fig:cone} that the eikonal
differential cross section changes slope at about $p_m$=250~MeV/c for
all values of $Q^2$ considered.  We remark that we imposed
quasi-elastic conditions for all cases contained in
Fig.~\ref{fig:cone}.  As a consequence, the momentum of the ejected
nucleon varies quite dramatically as one moves up in $Q^2$.  The
uniform behaviour of all curves contained in Fig.~\ref{fig:cone}
allows one to write down a relation between the transferred momentum
$\vec{q}$ and the polar scattering angle $\theta$ : $|\vec{q}| \,
\theta \leq 250$~MeV rad.  This simple relation could serve as a
conservative guideline to determine the opening angle of the cone in
which the outgoing proton momentum has to reside to ascertain that the
eikonal approximation produces ``realistic'' results.  This limitation
of the eikonal method can also be inferred from the results contained
in Refs.\ \cite{radici2,radici1}.  Indeed, in Figs.\ 3 and 4 of Ref.\
\cite{radici2} one can confirm that the above relation between
$|\vec{q}|$ and $\theta$ defines the missing momentum at which a
sudden change in the $p_m$ dependence of the calculated cross sections
is observed.  The above relation can be understood as follows. In
quasi-perpendicular and quasi-elastic kinematics, the missing momentum
roughly equals the transverse momentum of the ejected nucleon. With
increasing momentum transfer, the longitudinal momentum of the
escaping nucleon increases correspondingly while its transverse
momentum has to stay smaller than the suggested value of 250
MeV/c. Hence, the sine of the angle between the transferred momentum
and the ejectile's momentum has to decrease.  Since we are dealing
with small angles, sin($\theta$) can be approximated by $\theta$. The
opening angle of the cone in which the eikonal approximation is valid,
can be inferred to be independent of $Q^{2}$ in the Lorentz frame
where the ejected nucleon is at rest. When transforming back to the
labframe, lateral dimensions become dilated, and, thus, angles
contracted.

\subsection{The $Q^2$ evolution of off-shell effects}
\label{subsec:current}

A major point of concern in any A$(e,e'p)$B calculation are the
ambiguities regarding the off-shell electron-proton coupling.  Most
calculations do not obey current conservation and a variety of
prescriptions have been proposed to partially cure this
deficiency.  Here we adopt a heuristic view and estimate the
sensitivity of the calculated observables by comparing the results
obtained with different viable prescriptions for the electron-proton
coupling. Amongst the infinite number of possible prescriptions for the
off-shell electron-proton coupling we have selected four that are
frequently used in literature. Figure~\ref{fig:gaostruc12} shows the
separated structure functions for $1p_{1/2}$ knockout in the
kinematics of Fig.~\ref{fig:gaofig}.  Current conservation was imposed
by either modifying the longitudinal component of the vector current
operator (hereafter denoted as the ``J0 method''), or by modifying the
charge operator (hereafter denoted as the ``J3 method''), along the
lines of Eqs.\ (\ref{eq:long}) and (\ref{eq:charge}).  Note that for
the operator of Eq.\ (\ref{eq:donn}), both methods yield the same
results, since, by construction, this operator is current conserving,
regardless of the method adopted to compute the wave function for the
initial and final state.

Turning to the results shown in Fig.\ \ref{fig:gaostruc12}, the
predicted strengths in the longitudinal structure functions $R_{L}$
and $R_{TL}$ depend heavily on the choice made for the electron-proton
coupling.  For the $CC1$ prescription, for example, the values
obtained with the J3 method are several times bigger than those
obtained within the J0 method.  The predicted differences among the
various current operators within one scheme (``J0'' or ``J3'') are
also sizeable.  The ambiguities are, however, much smaller for the
calculations performed with the $J_{z} \rightarrow \frac{\omega}{q}
J_{0}$ substitution.  This clearly speaks in favor of this recipe
which is mostly used in A$(e,e'p)$ calculations. The $R_{TT}$ and
$R_T$ structure functions are, obviously, insensitive to whether the
``J0'' or ``J3'' method is adopted.  All adopted electron-proton
couplings but the $CC1$ one produce the same results in the $R_T$
and $R_{TT}$ responses.

With increasing $Q^{2}$ and the corresponding decreasing distance
scale, the role of off-shell ambiguities in the photon-nucleus coupling is
expected to decline and the impulse approximation is believed to
become increasingly accurate.  In order to investigate the degree and
rate to which this virtue may be realized, we have performed
calculations for kinematics in the range of $0.8 \leq Q^{2} \leq 20 \,
$(GeV/c)$^{2}$.  We use two techniques to estimate the relative
importance of the off-shell effects as a function of $Q^2$.  First,
results computed with the ``J0'' and ``J3'' method can be compared.
Second, predictions with various choices for the electron-proton
coupling are confronted with one another.  The validity of the IA is
then established whenever the final result happens to become
independent of the adopted choice.  In order to assess the degree to
which this independence is realized, we have considered ratios of
structure functions for some fixed kinematics but calculated with
different choices for the electron-proton coupling.  As a benchmark
calculation, we have computed $^{12}$C$(e,e'p)^{11}$B$(1p_{3/2}^{-1})$
observables in quasi-elastic kinematics for several values of the
four-momentum transfer.  The results are shown in Figs.\
\ref{fig:curr03} and \ref{fig:curr12}.  Fig.\
\ref{fig:curr03} shows for several observables the ratio of the values
obtained with the ``J3'' scheme to the corresponding prediction using
the ``J0'' scheme. Fig.\ \ref{fig:curr12} shows the ratio of the
strengths obtained with the $CC1$ vertex function compared to the
corresponding predictions with the $CC2$ form.  Remark that in the
limit of vanishing off-shell effects, these ratios should equal one.
It is indeed found that the calculations that are based on the
substitution $J_{z} \rightarrow \frac{\omega}{q} J_{0}$, tend to
converge to those based on $J_{0} \rightarrow \frac{q}{\omega} J_{z}$
with increasing energy transfer.  This is particularly the case at low
missing momenta, where the decrease in the longitudinal response is
almost exponential.  The overall behaviour is identical for the higher
missing momentum case ($p_{m}$ = 150~MeV/c), but the rate of decrease
is somewhat slower. This can be attributed to the fact that at higher
momenta, hence, greater angles, the transverse components of the
vertex functions play a more important role.  Looking at Fig.\
\ref{fig:curr12} one can essentially draw the same conclusions.  The
predictions with the different prescriptions also converge to each
other as the energy is increased. Again this convergence is more
pronounced for the low missing momentum case.  This feature is most
apparent in the purely transverse channel, which dominates the cross
section at sufficiently high energies.  It appears thus as if
off-shell ambiguities, speaking in terms of strenghts and absolute
cross sections, are of far less concern at higher $Q^{2}$ than they
used to be in the $Q^{2} \leq$ 1~(GeV/c)$^{2}$ region, where most of
the data have been accumulated up to now.  The interference structure functions
$R_{TT}$ and $R_{TL}$ are subject to off-shell ambiguities that are
apparently extending to the highest four-momentum transfers considered
here. This feature was already established in Ref.\ \cite{rpwia}
and explained by referring to the large weight of the negative energy
solutions in the interference structure functions $R_{TL}$ and
$R_{TT}$.  

\subsection{Relativistic Effects}
\label{subsec:relativity}

Recently, there have been several claims for strong indications for
genuine (or, ``dynamic'') relativistic effects in
A$(\vec{e},e'\vec{p})$ observables
\cite{gao,gardner94,udias99,joha99}.  In an attempt to implement some
of these effects in calculations based on a Schr\"{o}dinger picture,
several techniques to obtain a ``relativized version'' of the
electron-nucleus vertex have been developed.  In leading order in a
$p/M$ expansion these ``relativized'' electron-nucleus vertices
typically miss the coupling between the lower components in the bound
and scattering states. For that reason, we interpret the effect of the
coupling between the lower components in the bound and scattering
states as a measure for the importance of relativistic effects.  In
Fig.\ \ref{fig:relativity} we display results of fully relativistic
$^{12}$C$(e,e'p)^{11}$B$(1p_{3/2}^{-1})$ calculations and calculations
in which the specific coupling between the lower components in the
bound and scattering states have been left out.  We consider
quasi-elastic conditions and study the $Q^2$ evolution of the
structure functions for two values of the missing momentum ($p_{m}$ = 0 and 150
MeV/c) both corresponding with small recoil angles.  Hence, the
results of Fig.~\ref{fig:relativity} refer to kinematic conditions for
which the eikonal approximation is justified. A rather complex and
oscillatory $Q^2$ dependence of the relativistic effects emerges from
our numerical calculations.    
Looking first at the $p_m \approx$ 0~MeV/c case, which nearly
corresponds with parallel kinematics, we observe that for both the
longitudinal and transverse structure functions, the impact of the
coupling amongst the lower components first increases, and then tends
to become fairly constant for higher values of $\omega$.  The
genuine relativistic effect stemming from the coupling between the
lower components in the initial and final states is larger in the
longitudinal than in the transverse channel.  It is noteworthy that in
the cross section the impact of the ``relativistic dynamical effects'' never
exceeds the 10 \% level.  If we turn our attention to the interference
structure functions $R_{TL}$ and $R_{TT}$, the relativistic effects
grow in importance.  Especially for the $R_{TL}$ structure function
the effects are large and extend to the smallest values of $Q^2$
considered here.  This enhanced sensitivity of the $R_{TL}$ response
to relativistic effects, even when relatively low values of $Q^2$ are
probed, complies with the conclusions drawn in other studies
\cite{kellypol,gardner94,janmeson,ulrych,gilad}.  Also the tendency of the
relativistic effects to increase the cross section when higher values
of $p_m$ are probed complies with the findings of earlier studies
\cite{udias96}.  A quantity that is relatively easy to access
experimentally and depends heavily upon the $R_{TL}$
term, is the so-called left-right asymmetry $A_{LT}$
\begin{eqnarray}
A_{LT} = \frac{\sigma (\phi = 0^{\circ}) - \sigma (\phi =
180^{\circ})}{\sigma (\phi = 0^{\circ}) + \sigma (\phi =
180^{\circ})} = \frac{v_{TL} R_{TL}} {v_L R_L + v_T R_T + v_{TT}
R_{TT}} \; .
\end{eqnarray}
In Fig.\ \ref{fig:asymmetry} we have plotted the left-right asymmetry
for both $1p_{1/2}$ and $1p_{3/2}$ knockout from $^{16}$O in the
kinematics of Fig.\ \protect\ref{fig:gaofig}. It is indeed verified
that the asymmetry is very sensitive to relativistic effects.  As has
been reported, relativistic effects enhance the asymmetry further, and
this enhancement is more pronounced for the $1p_{1/2}$ knockout
reaction. The role played by the lower components in this dynamical
enhancement of the left-right asymmetry can be further clearified by
looking at the results of Fig.\ \ref{fig:Q2asym}. In this figure, we
plot the left-right asymmetry for $1p_{3/2}$ knockout from $^{12}$C,
for different $Q^{2}$ and quasi-elastic conditions. Looking at the
fully relativistic curves, we observe a gradual decrease of the
asymmetry with increasing $Q^{2}$.  At the same time, the relative
contribution of the ``non-relativistic'' contribution to $A_{LT}$
diminishes.  This feauture indicates that the asymmetry $A_{LT}$ is
nearly exclusively generated by the coupling between the lower
components as $Q^2$ increases.

\section{Summary}
\label{sec:conclusion}

We have outlined a fully relativistic eikonal framework for
calculating cross sections for $(e,e'p)$ reactions from spherical
nuclei at intermediate and high four-momentum transfers and carried
out $^{12}$C$(e,e'p)$ and $^{16}$O$(e,e'p)$ calculations for a variety
of kinematical conditions, thereby covering four-momentum transfers in
the range $0.8 \leq Q^{2} \leq 20 \ $ (GeV/c)$^{2}$.  Our results
illustrate that the validity of the eikonal method is
confined to proton emission in a cone with a relatively small opening
angle about the direction of the virtual photon's
momentum. This observation puts serious constraints on the
applicability of the Glauber method, that is based on the eikonal
approximation, for modelling the final-state interactions in
high-energy $(e,e'p)$ reactions from nuclei. Incorporation of the
inelastic channels in the eikonal method is however needed to fully
appreciate the limits of the Glauber model, and work along these lines
is in progress.  In line with the expectations, our investigations
illustrate that the uncertainties induced by off-shell ambiguities on
the calculated observables diminish as $Q^2$ increases.  Nevertheless,
in the relativistic eikonal framework four-momentum transfers of the
order 5~(GeV/c)$^2$ appear necessary to assure that the effect of the
off-shell ambiguities can be brought down to the percent level.  Our
theoretical framework permits to assess the impact of the relativistic
effects over a wide energy range.  The impact of the lower components
on the $(e,e'p)$ observables is observed to be significant over the
whole $Q^2$ range studied.  Especially the left-right asymmetry lends
itself very well to study these effects of genuine relativistic
origin.

{\bf Acknowledgement}\\
This work was supported by the Fund for Scientific Research of
Flanders under contract No.\ 4.0061.99 and the University Research
Council.

\newpage

\begin{figure}[tbp]
\caption{The calculated scalar ($\rho_{s}$), baryon ($\rho_{B}$), rho
($\rho_{3}$) and proton ($\rho_{P}$) density distributions in $^{12}$C
and $^{16}$O.}  
\label{fig:densities}
\end{figure}

\begin{figure}[tbp]
\caption{The radial dependence of the scalar and vector potentials (in
absolute values) as obtained from relativistic Hartree calculations
for $^{12}$C and $^{16}$O.}
\label{fig:potentials}
\end{figure} 

\begin{figure}[tbp]
\caption{Measured $^{16}$O$(e,e'p)$ cross sections
compared to relativistic eikonal and RPWIA calculations at $\epsilon$
= 2.4~GeV, $q$ = 1~GeV/c and $\omega$ = 0.439~GeV in quasi-perpendicular
kinematics.  The calculations use the current operator $CC1$.  The data
are from Ref.\ \protect\cite{gao}.}
\label{fig:gaofig}
\end{figure}

\begin{figure}[tbp]
\caption{The reduced cross section for the
$^{16}$O$(e,e'p)^{15}$N$(1p_{3/2}^{-1})$ reaction versus missing momentum
at three values of the recoil angle $\theta$. A fixed outgoing proton
momentum of $|\vec{p}| = 1$~GeV was considered.  The solid line shows the
fully relativistic eikonal calculation, while the dashed one shows the
RPWIA results. The calculations use the $CC1$ prescription.}
\label{fig:fixtp}
\end{figure}

\begin{figure}[tbp]
\caption{The differential cross section for the
$^{12}$C$(e,e'p)^{11}$B$(1p_{3/2}^{-1})$ reaction versus missing
momentum at six different values for $Q^{2}$.  Quasi-elastic
conditions and perpendicular kinematics were considered.}
\label{fig:cone}
\end{figure}

\begin{figure}[tbp]
\caption{The different structure functions versus missing momentum for
$1p_{1/2}$ knockout from $^{16}$O in the kinematics of Fig.\
\protect\ref{fig:gaofig}.  The calculations in the left column imposed
current conservation by replacing the longitudinal component of the
vector current operator (Eq.\ \protect\ref{eq:long}), while for the
results in the right column the charge density operator was modified
according to Eq.~(\protect\ref{eq:charge}).  The curves refer to the
different off-shell prescriptions as they were introduced in
Sect.~\ref{subsec:vertex}.}
\label{fig:gaostruc12}
\end{figure}

\begin{figure}[tbp]
\caption{The $Q^{2}$ dependence of the sensitivity of the calculated
$(e,e'p)$ structure functions to the choice for the 
electron-nucleus vertex for $1p_{3/2}$ knockout from $^{12}$C in
quasi-elastic kinematics.  The curves show for the various observables
the ratio of the predictions with the ``J3'' method to those obtained
with the ``J0'' method.  Solid (dashed) line corresponds with the $p_{m} = 0$~
MeV/c ($p_{m} = 150$~MeV/c) situation.}
\label{fig:curr03}
\end{figure}

\begin{figure}[tbp]
\caption{The $Q^2$ dependence of the sensitivity of the $(e,e'p)$ structure functions to
the choice for the electron-nucleus vertex for $1p_{3/2}$ knockout
from $^{12}$C.  The curves display the ratio of the predictions  using
the vertex function $\Gamma _{cc1}^{\mu}$ to those using $\Gamma
_{cc2}^{\mu}$.  Solid (dashed) lines correspond with $p_{m} = 0$~MeV/c
($p_{m} = 150$~MeV/c).}
\label{fig:curr12}
\end{figure}

\begin{figure}[tbp]
\caption{The  $Q^2$ dependence of the sensitivity of the $(e,e'p)$
structure functions to dynamical relativistic effects.  The curves
show for $1p_{3/2}$ knockout from $^{12}$C the ratio of the fully
relativistic results to the predictions when the coupling between the
lower components is neglected.  The solid (dashed) line presents
results for the situation $p_{m} = 0$~MeV/c ($p_{m} = 150$~MeV/c).}
\label{fig:relativity}
\end{figure}

\begin{figure}[tbp]
\caption{The left-right asymmetry $A_{LT}$ for both $1p_{1/2}$ and $1p_{3/2}$
knockout from $^{16}$O in the kinematics of Fig.\ 
\protect\ref{fig:gaofig}. The data are from Ref.~\protect \cite{gao}.} 
\label{fig:asymmetry}
\end{figure}

\begin{figure}[tbp]
\caption{The left-right asymmetry $A_{LT}$ for $1p_{3/2}$ knockout
from $^{12}$C for different $Q^{2}$, under quasi-elastic conditions
and perpendicular kinematics.} 
\label{fig:Q2asym}
\end{figure}

\newpage

\begin{figure}
\begin{center}
{\mbox{\epsfysize=10cm\epsffile{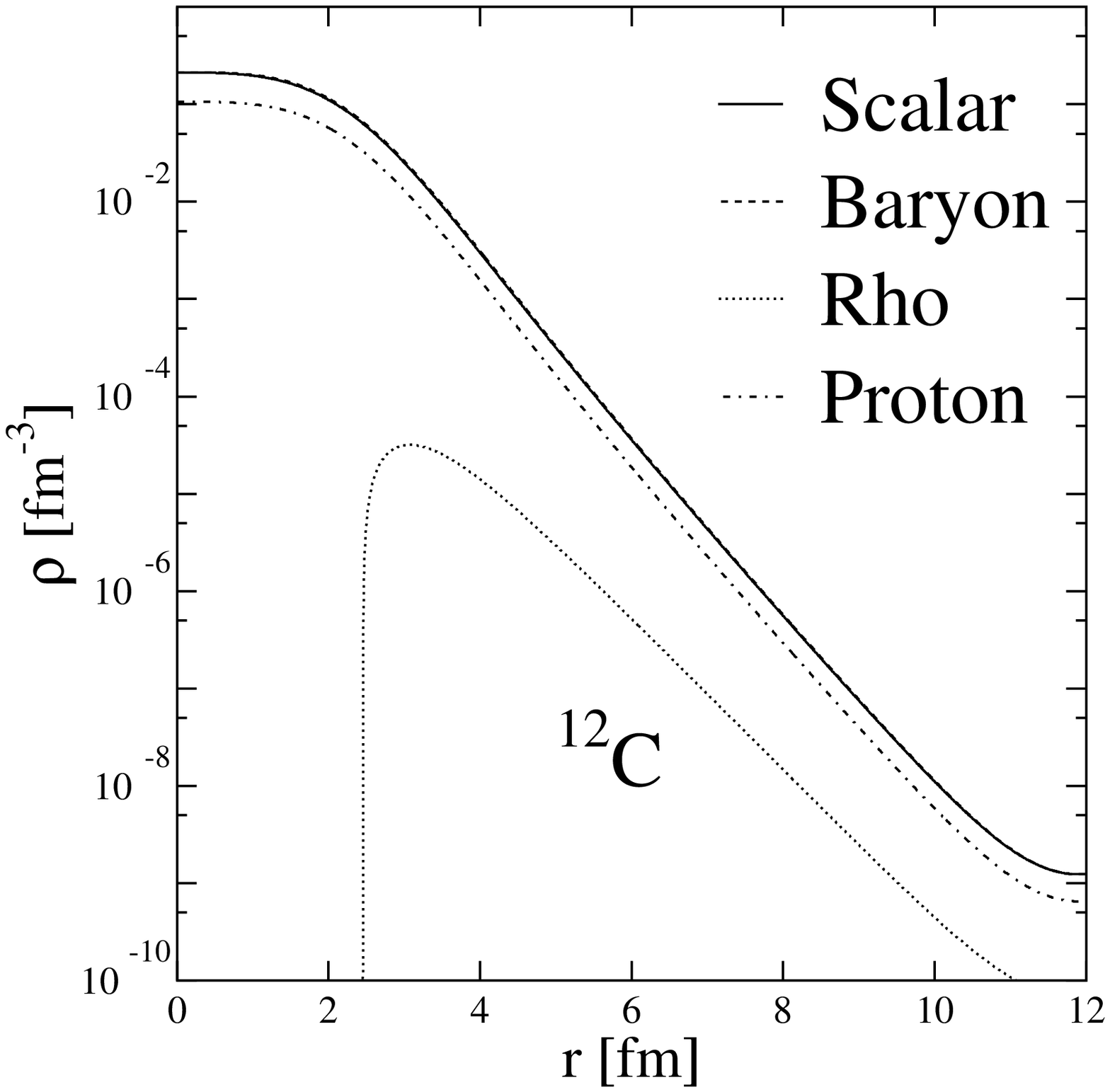}}}
{\mbox{\epsfysize=10cm\epsffile{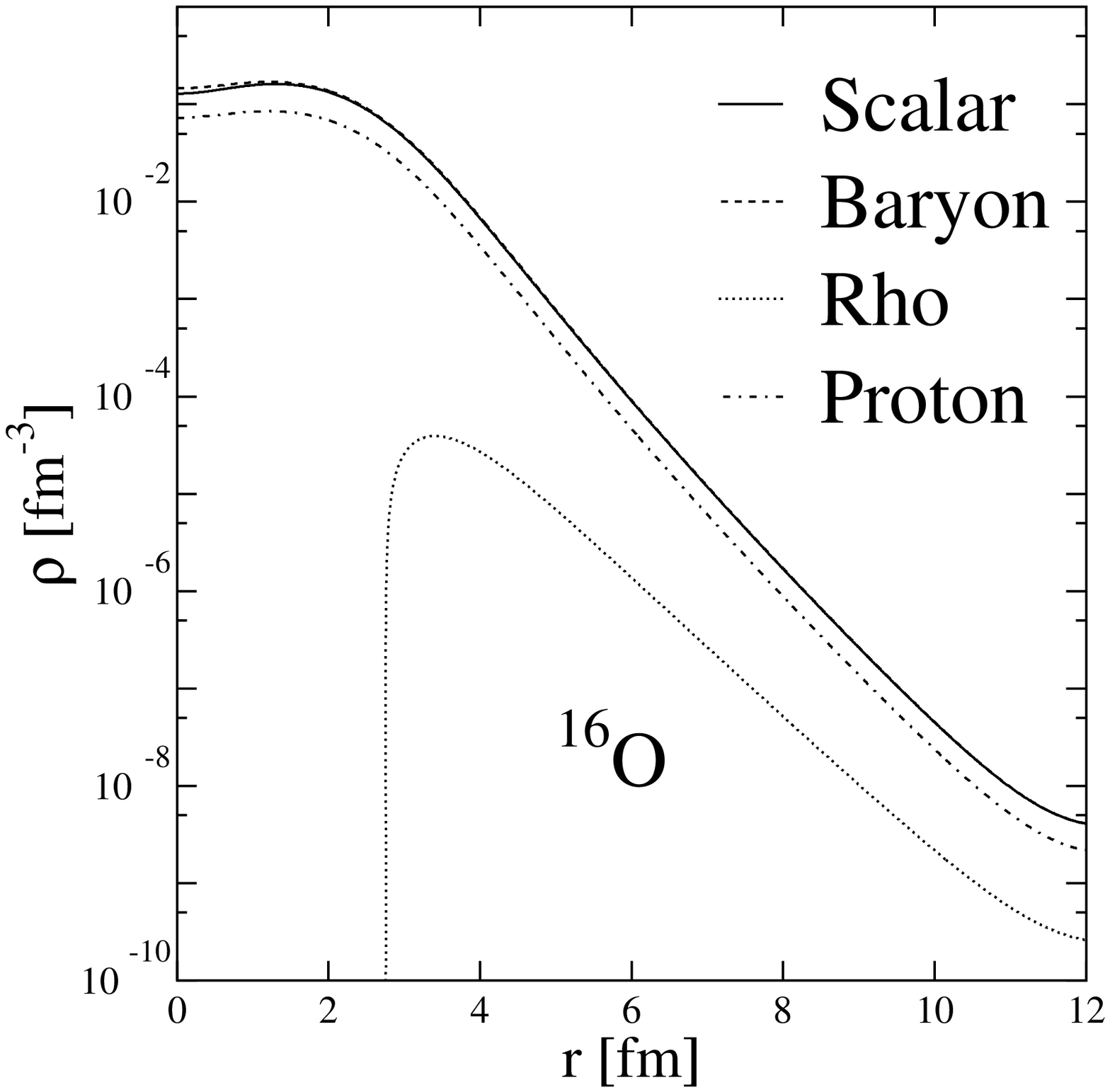}}}
\end{center}
\end{figure}

\begin{center}
{\Huge Figure 1}
\end{center}

\newpage

\begin{figure}
\begin{center}
{\mbox{\epsfysize=12.cm\epsffile{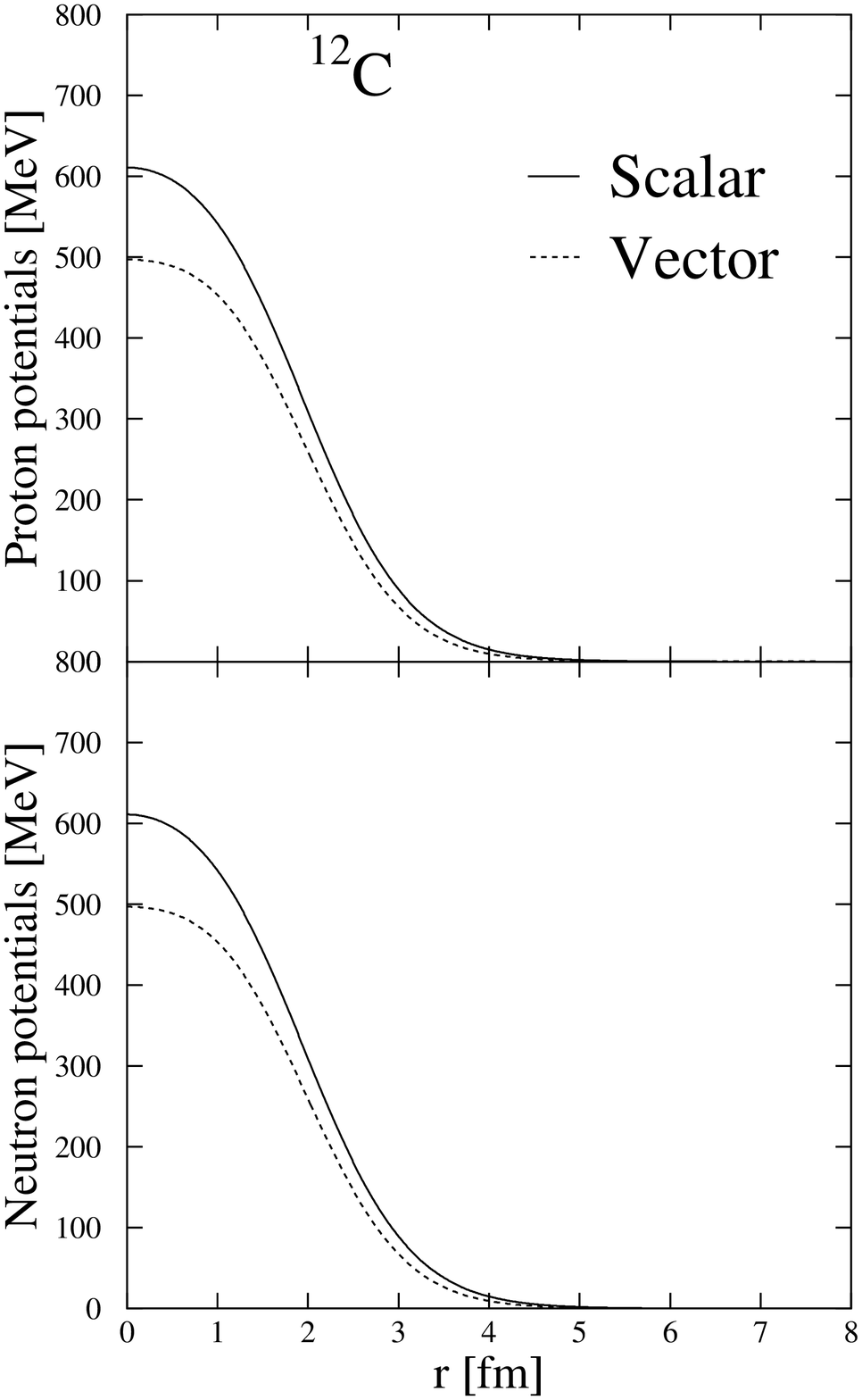}}}
{\mbox{\epsfysize=12.cm\epsffile{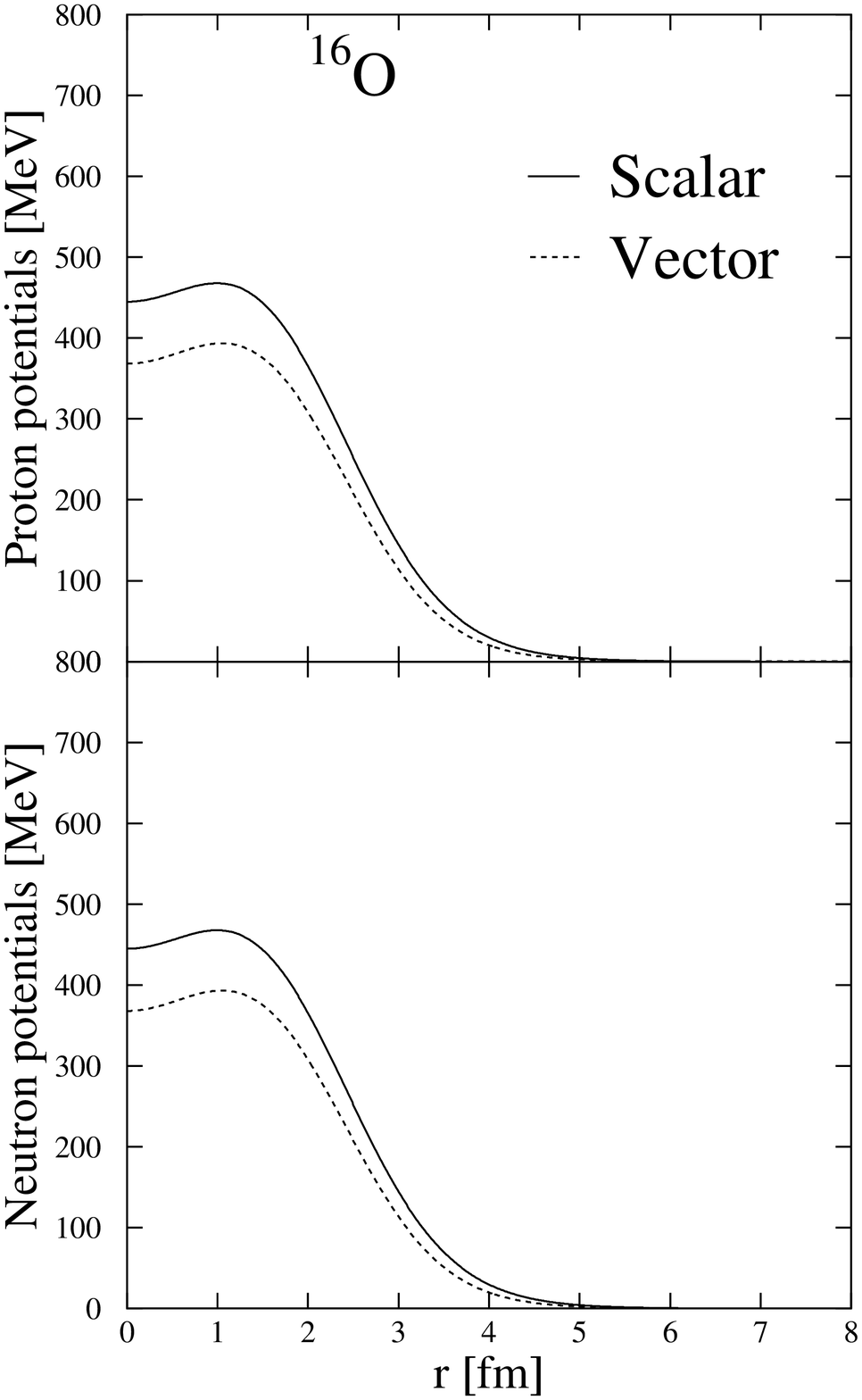}}}
\end{center}
\end{figure} 

\begin{center}
{\Huge Figure 2}
\end{center}

\newpage

\begin{figure}
\begin{center}
{\mbox{\epsfysize=15cm\epsffile{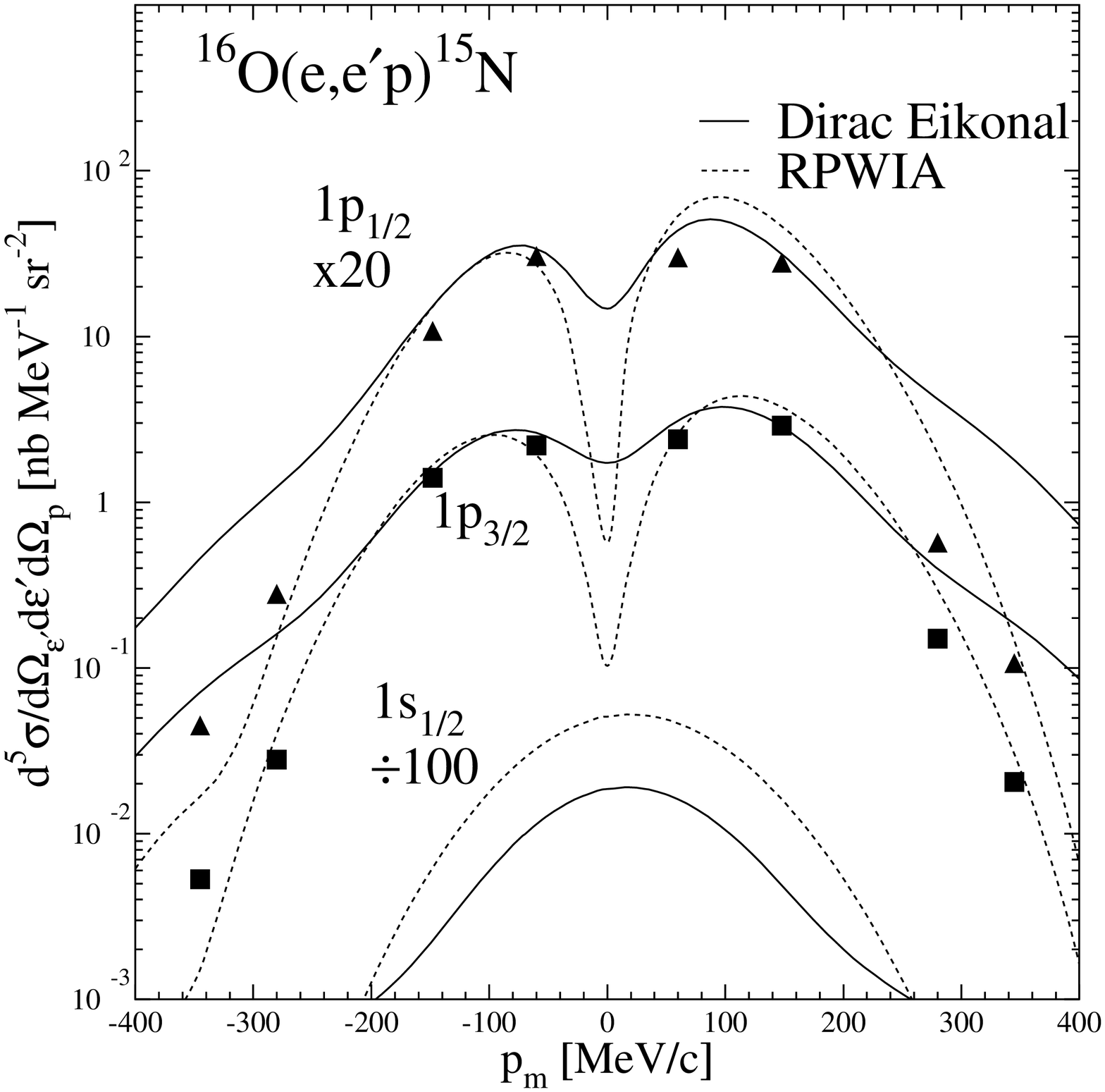}}}
\end{center}
\end{figure}

\begin{center}
{\Huge Figure 3}
\end{center}

\newpage

\begin{figure}
\begin{center}
{\mbox{\epsfysize=15cm\epsffile{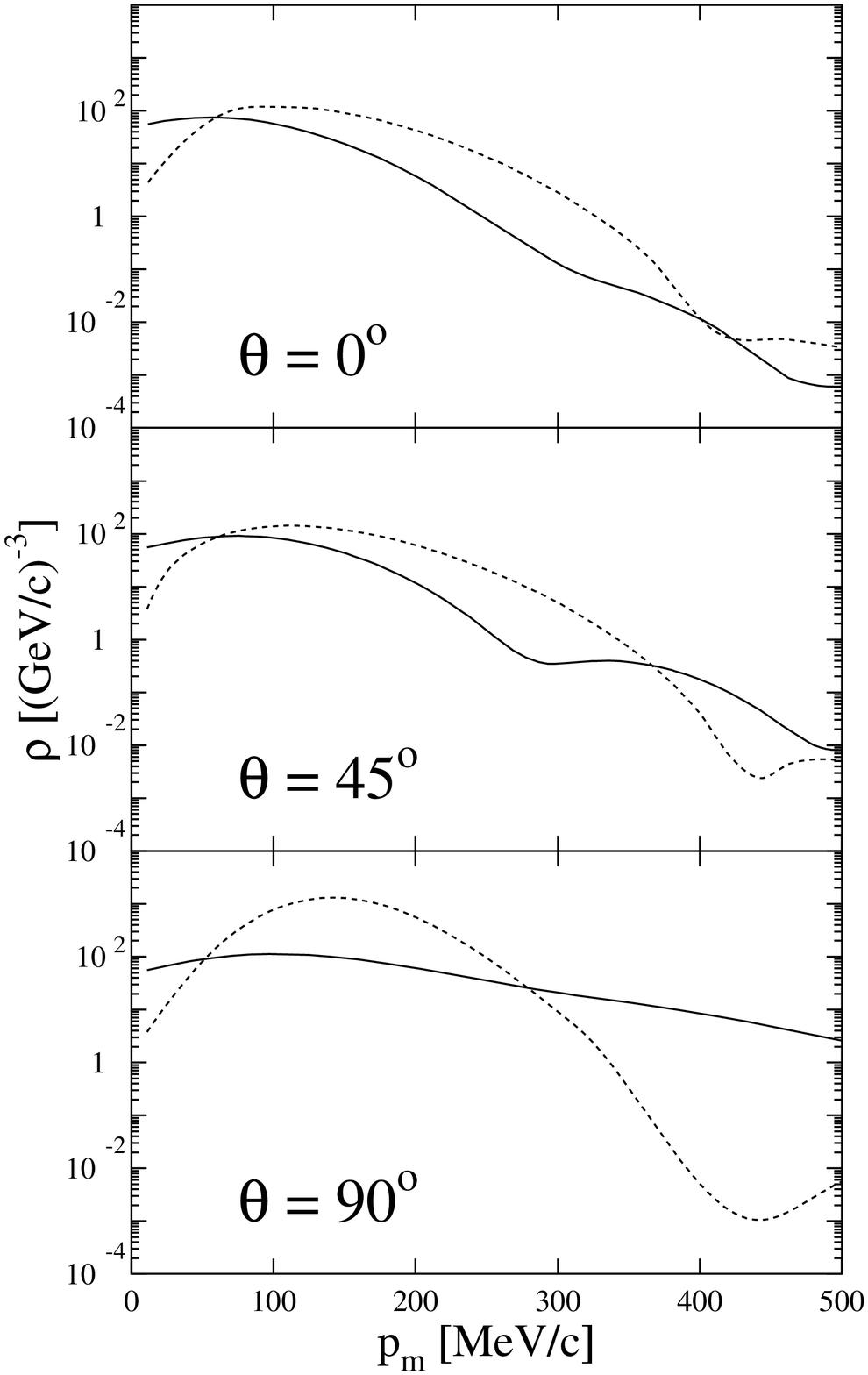}}}
\end{center}
\end{figure}

\begin{center}
{\Huge Figure 4}
\end{center}

\newpage

\begin{figure}
\begin{center}
{\mbox{\epsfysize=17.cm\epsffile{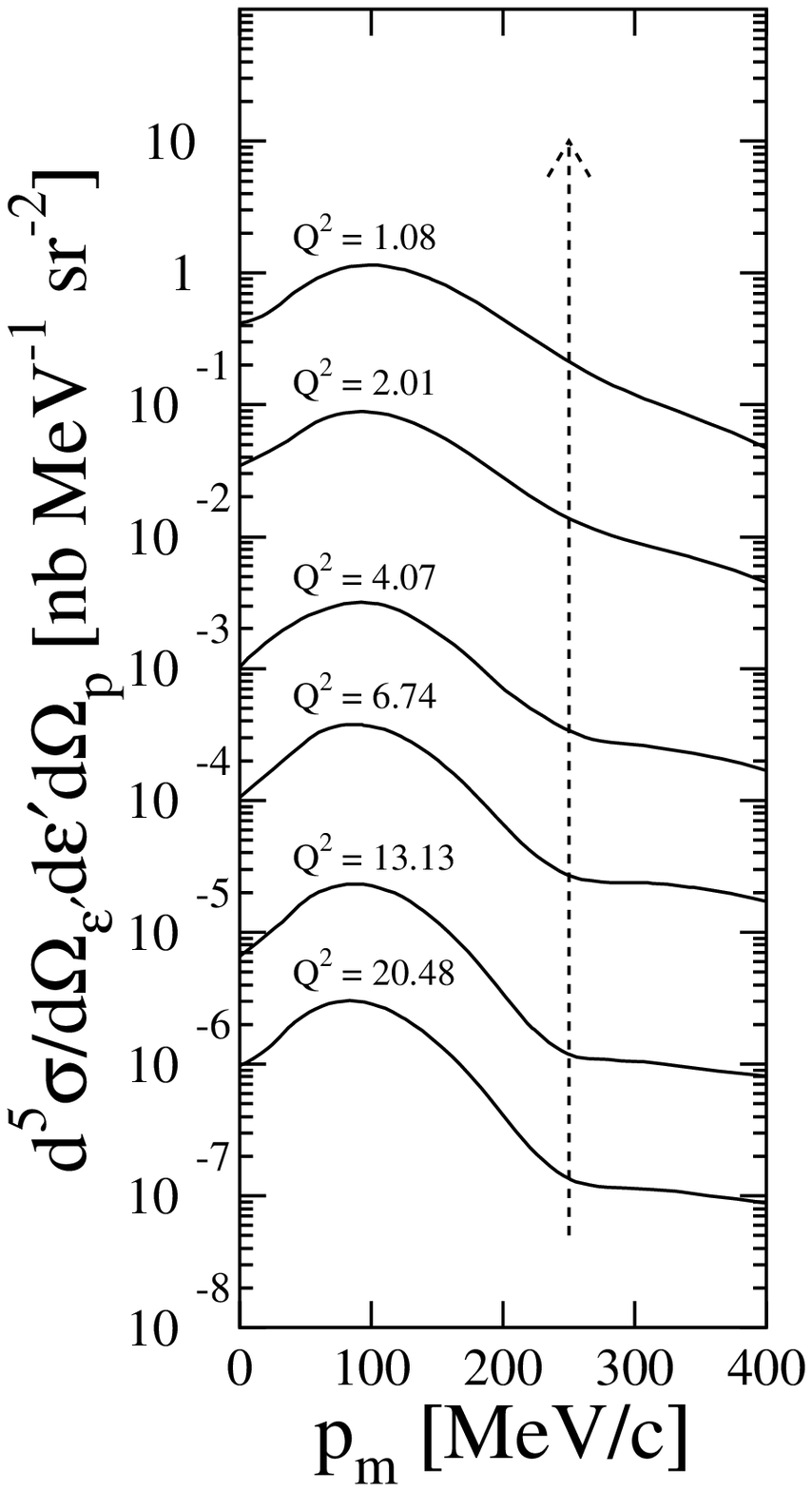}}}
\end{center}
\end{figure}

\begin{center}
{\Huge Figure 5}
\end{center}

\newpage

\begin{figure}
\begin{center}
{\mbox{\epsfysize=17.cm\epsffile{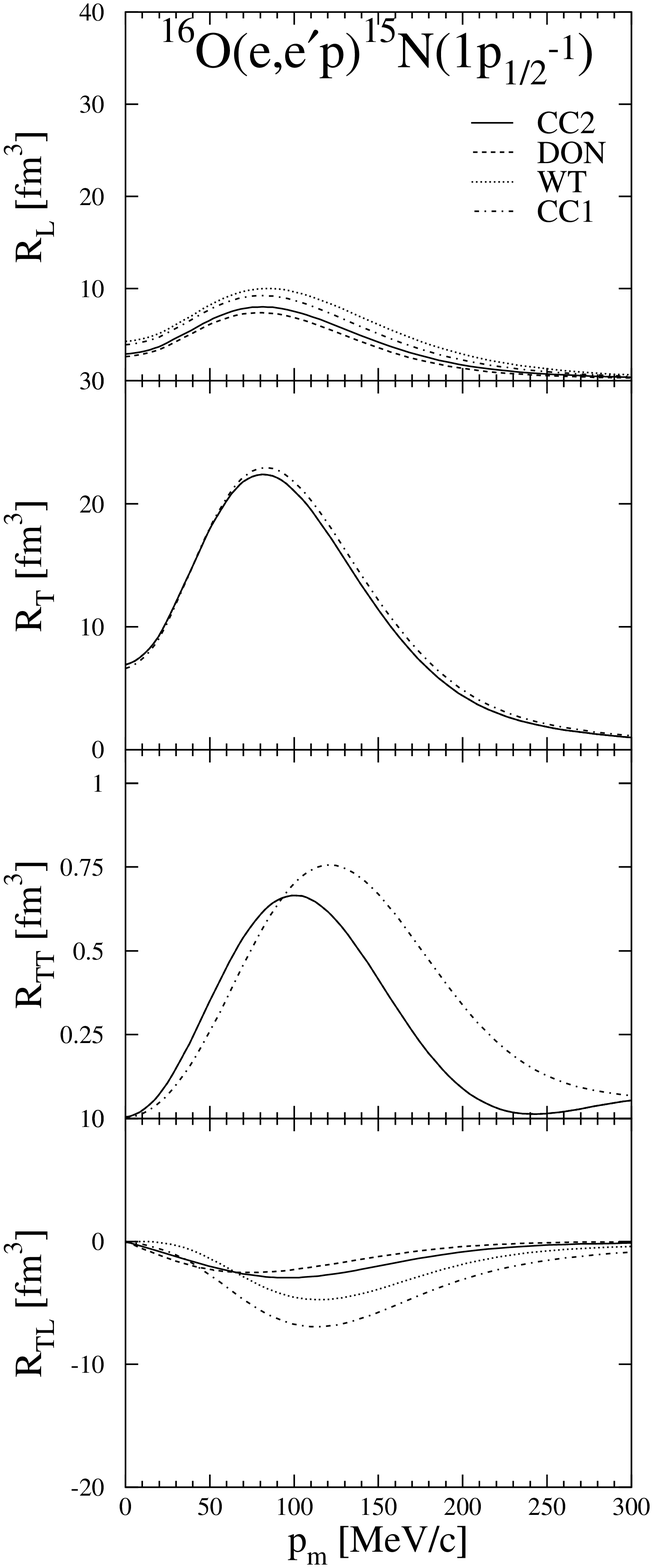}}}
{\mbox{\epsfysize=17.cm\epsffile{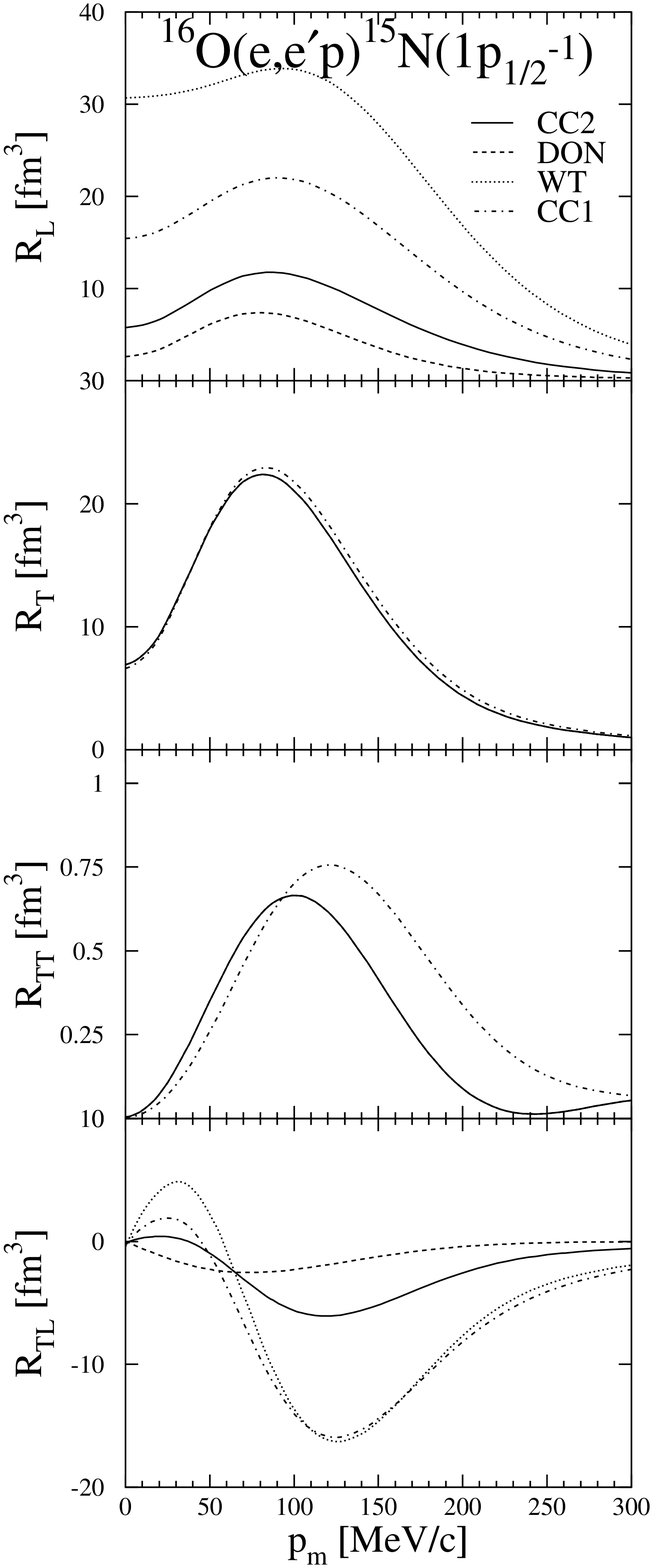}}}
\end{center}
\end{figure}

\begin{center}
{\Huge Figure 6}
\end{center}

\newpage

\begin{figure}
\begin{center}
{\mbox{\epsfysize=18.cm\epsffile{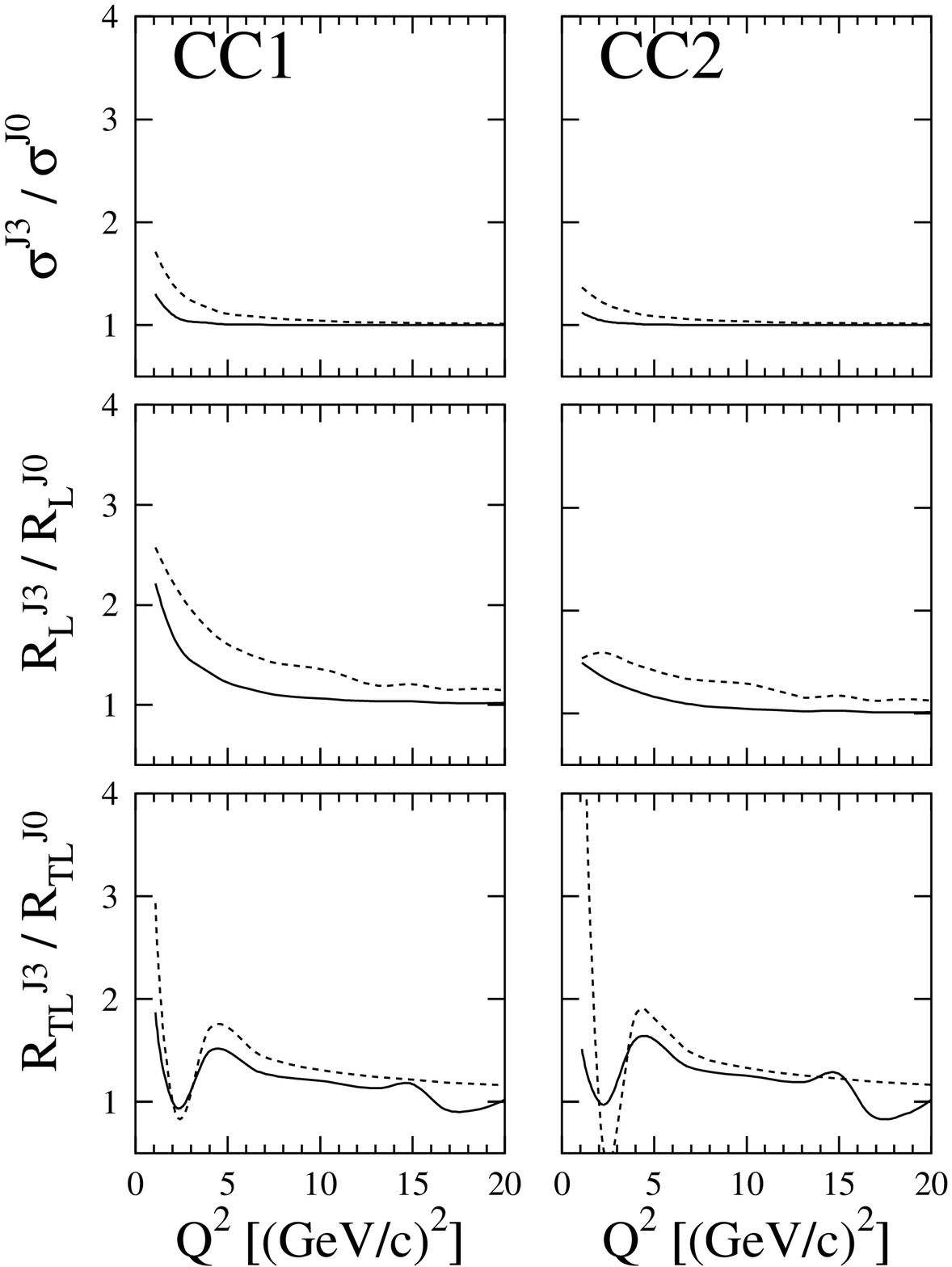}}}
\end{center}
\end{figure}

\begin{center}
{\Huge Figure 7}
\end{center}

\newpage

\begin{figure}
\begin{center}
{\mbox{\epsfysize=18.cm\epsffile{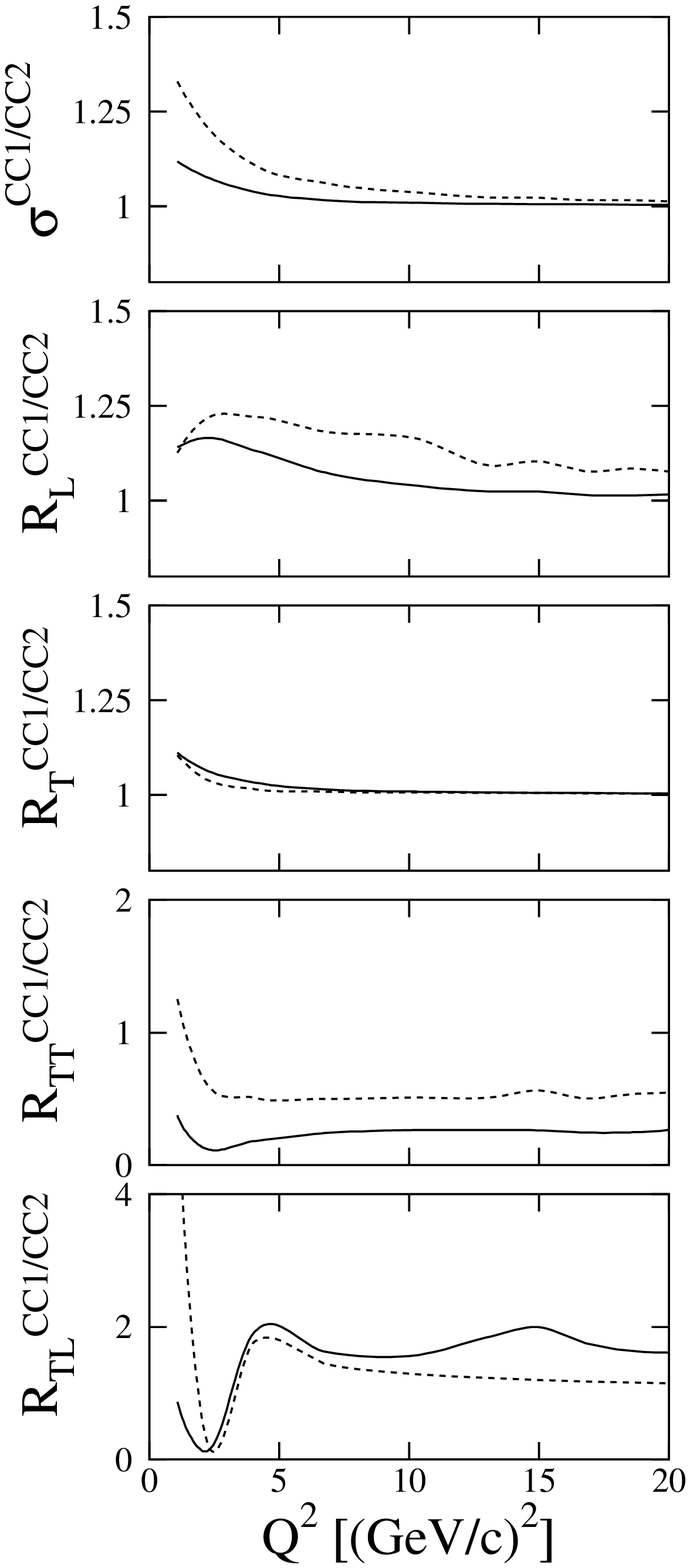}}}
\end{center}
\end{figure}

\begin{center}
{\Huge Figure 8}
\end{center}

\newpage

\begin{figure}
\begin{center}
{\mbox{\epsfysize=18.cm\epsffile{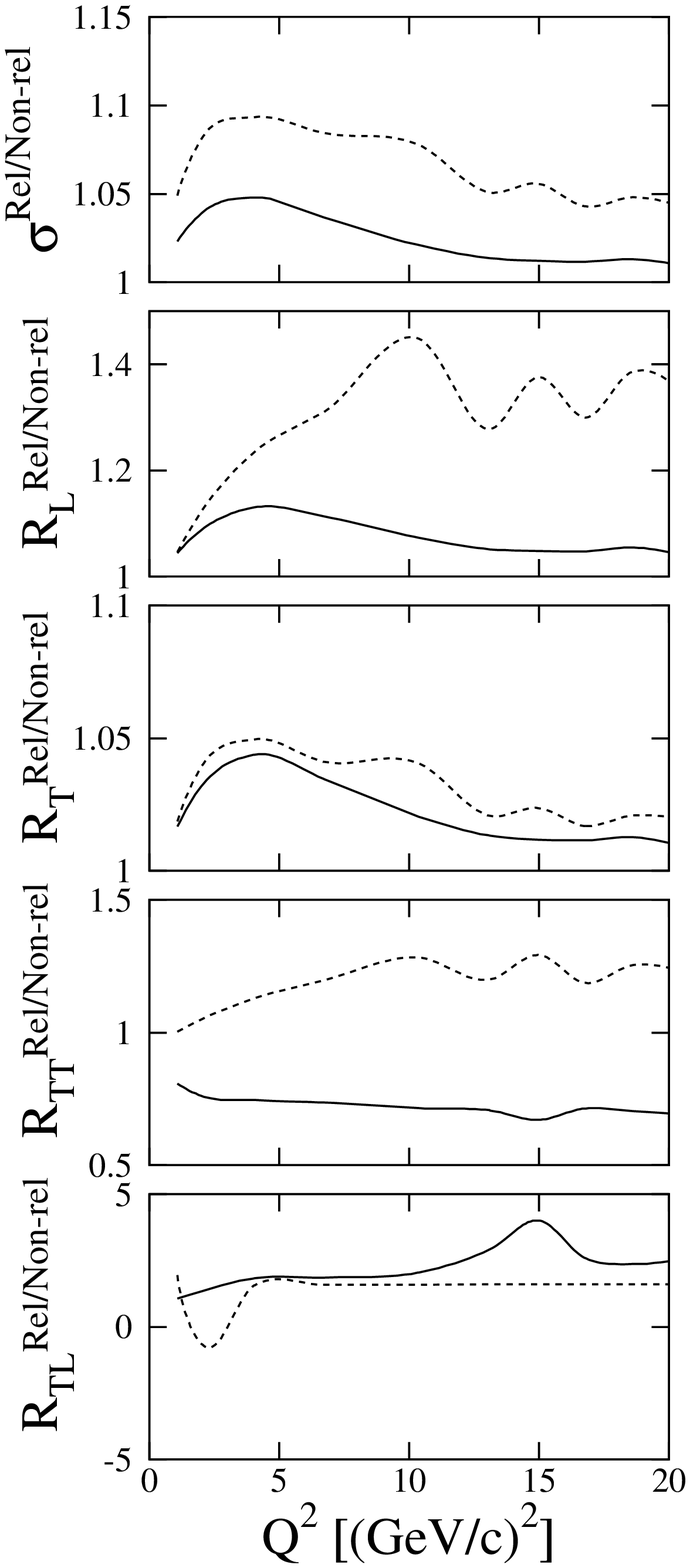}}}
\end{center}
\end{figure}

\begin{center}
{\Huge Figure 9}
\end{center}

\newpage

\begin{figure}
\begin{center}
{\mbox{\epsfysize=12.cm\epsffile{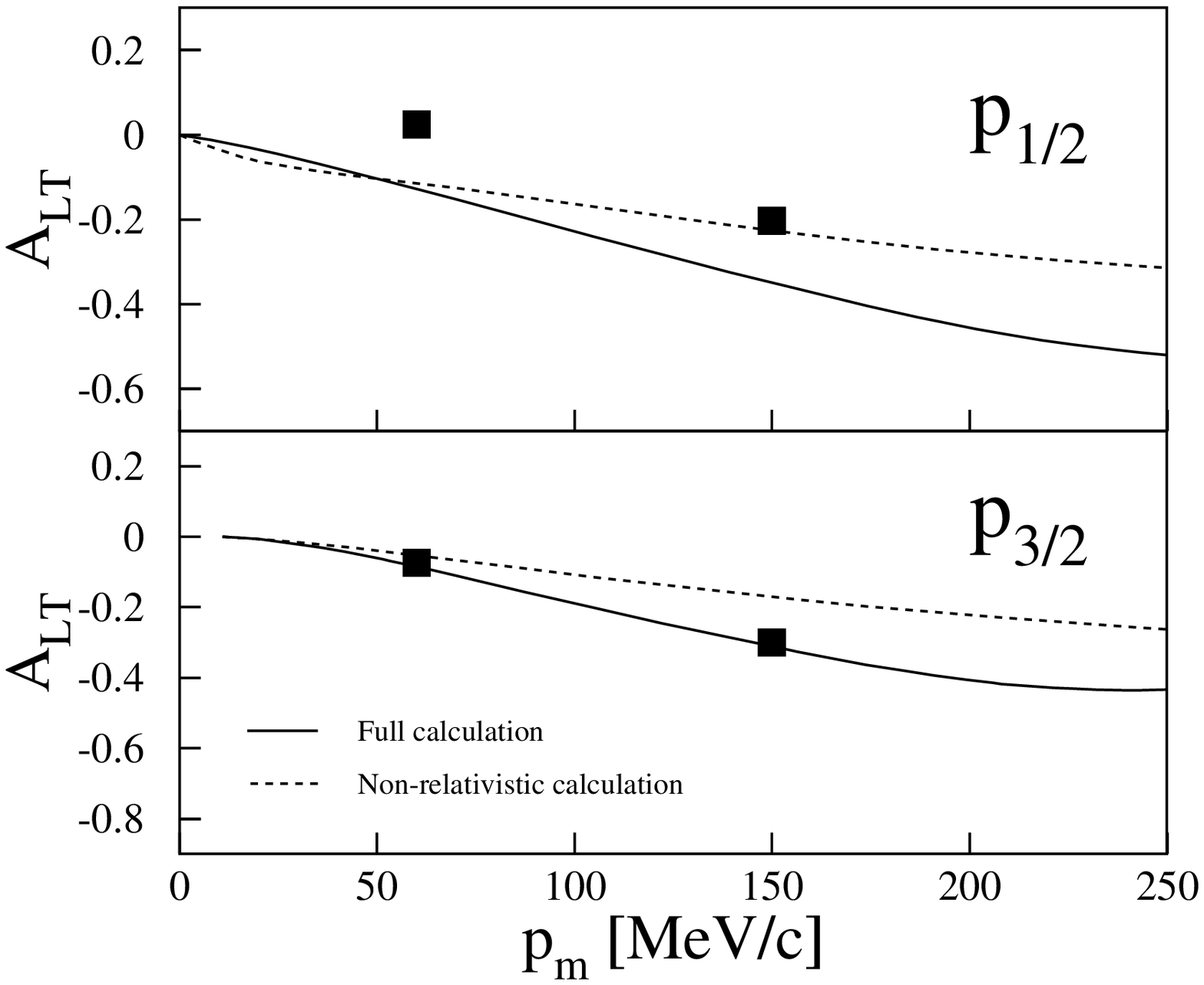}}}
\end{center}
\end{figure}

\begin{center}
{\Huge Figure 10}
\end{center}

\newpage

\begin{figure}
\begin{center}
{\mbox{\epsfysize=12.cm\epsffile{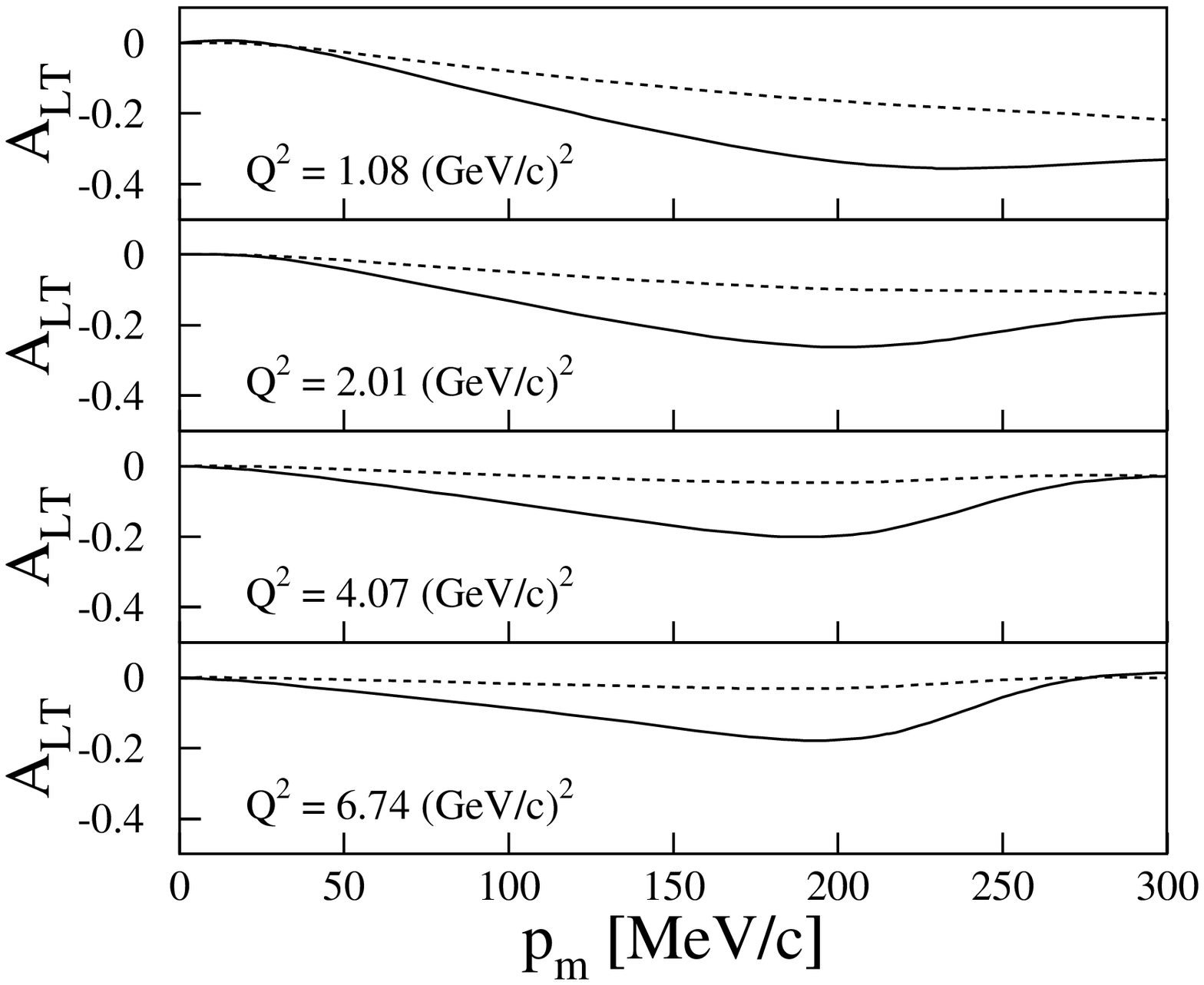}}}
\end{center}
\end{figure}

\begin{center}
{\Huge Figure 11}
\end{center}

\end{document}